\documentclass[11pt,letterpaper,english,aps]{revtex4}
\usepackage{times}
\usepackage[T1]{fontenc}
\usepackage[latin1]{inputenc}
\usepackage{array}
\usepackage{verbatim}
\usepackage{subfigure}
\usepackage{float}
\usepackage{amsmath}
\usepackage{graphicx}

\makeatletter

\providecommand{\tabularnewline}{\\}

\makeatother
\begin{document}

\title{Investigation of pulsed laser induced dewetting in nanoscopic metal
films}

\author{$^{\text{1,2}}$Justin Trice, $^{3}$Dennis Thomas, $^{\text{1,2}}$Christopher
Favazza, $^{\text{2,3}}$Radhakrishna Sureshkumar%
\thanks{suresh@che.wustl.edu%
}, and $^{\text{1,2}}$Ramki Kalyanaraman%
\thanks{contact author at ramkik@wuphys.wustl.edu%
}}

\affiliation{$^{\text{1}}$Department of Physics, Washington University in St.
Louis, MO 63130}

\affiliation{$^{\text{2}}$Center for Materials Innovation, Washington University
in St. Louis, MO 63130}

\affiliation{$^{\text{3}}$Department of Energy, Environmental and Chemical Engineering,
Washington University in St. Louis, MO 63130\thispagestyle{empty}}

\begin{abstract}
Hydrodynamic pattern formation (PF) and dewetting resulting from pulsed
laser induced melting of nanoscopic metal films have been used to
create spatially ordered metal nanoparticle arrays with  monomodal
size distribution on SiO$_{\text{2}}$/Si substrates. PF was investigated
for film thickness $h\leq7\, nm$ < laser absorption depth $\sim11\, nm$
and different sets of laser parameters, including energy density $E$
and the irradiation time, as measured by the number of pulses $n$.
PF was only observed to occur for $E\geq E_{m}$, where {\normalsize $E_{m}$}
denotes the $h$-dependent threshold energy required to melt the film.
\emph{Even at such small length scales}, theoretical predictions for
$E_{m}$ obtained from a continuum-level lumped parameter heat transfer
model for the film temperature, coupled with the 1-D transient heat
equation for the substrate phase, were consistent with experimental
observations provided that the thickness dependence of the reflectivity
of the metal-substrate bilayer was incorporated into the analysis.
The model also predicted that perturbations in \emph{h} would result
in intrinsic thermal gradients $\partial T/\partial h$ whose magnitude
and sign depend on $h$, with $\partial T/\partial h>0$ for $h<h_{c}$
and $\partial T/\partial h<0$ for $h>h_{c}\approx9\, nm$. For the
thickness range investigated here, the resulting thermocapillary effect
 was minimal since the thermal diffusion time $\tau_{H}\leq$ the
pulse time. Consequently, the spacing between the nanoparticles and
the particle diameter  were found to increase as $h^{2}$ and $h^{5/3}$
respectively, which is consistent with the predictions of the thin
film hydrodynamic (TFH) dewetting theory. PF  was characterized by
the appearance of discrete holes followed by bicontinuous or cellular
patterns which finally consolidated into nanoparticles via capillary
flow rather than via Rayleigh-like instabilities reported for low
temperature dewetting of viscous liquids. This difference is attributed
to the high capillary velocities of the liquid metal arising from
its relatively large interfacial tension and low viscosity as well
as the smaller length scales of the liquid bridges in the experiments.
 The predicted liquid phase lifetime $\tau_{L}$ was between $2-15\, ns$,
which is much smaller than the dewetting time $\tau_{D}\geq25\, ns$
as predicted by the linear TFH theory. Therefore, dewetting required
the application of multiple pulses. During the early stages of dewetting,
the ripening rate, as measured by the rate of change of characteristic
ordering length with respect to $n$, increased linearly with $E$
due to the linear increase in $\tau_{L}$ with increasing $E$ as
predicted by the thermal model. The final nanoparticle spacing was
 robust, independent of $E$ and $n$, and only dependent on $h$
due to the relatively weak temperature dependence of the thermophysical
properties of the metal (Co). These results suggest that fast thermal
processing combined with the unique thermophysical parameters of metals
can lead to novel pattern formation, including quenching of a wide
range of length scales and morphologies. 
\end{abstract}
\maketitle

\section{Introduction}

Physical phenomena that impose patterns with characteristic length
scales  can be utilized for the self-assembly of spatially ordered
metal nanoparticles which have various applications, including in
magnetics, catalysis, nanophotonics and plasmonics \cite{sun00,fan99,maier03,maier03b,quinten98,white00,garcia06c}.
Nonlinear dynamical instabilities which evolve into a robust stable
state characterized by a narrow range of length scales are especially
well-suited for this purpose. Typically such self-assembly of ordered
patterns on surfaces and thin films arises from the competition between
various physico-chemical effects. For example, surface rippling under
ion irradiation can be attributed to an instability resulting from
the competition between irradiation induced ion erosion  and smoothing
due to surface diffusion. The length and time scales of the ensuing
patterns are determined by ion flux, surface temperature and surface
diffusion parameters \cite{Sigmund65,harper88,Chason94,erlebacher99,facsko01,valbusa03,teichert04}\textbf{.}
Alternately, nanopatterns could result from instabilities of a thin
fluid layer leading to spatio-temporal re-organization of the fluid
with well defined  length and time scales that depend on the thermophysical
material properties such as interfacial tension, fluid/substrate contact
angle, fluid viscosity and, for ultrathin films, long range dispersion
forces such as van der Waal's interactions. An example of  pattern
formation by such instabilities is dewetting of thin films caused
by a hydrodynamic instability which occurs when attractive intermolecular
forces overcome the stabilizing effect of interfacial tension \cite{vrij66,sharma86,sharma93,Seemann01}.
Under such conditions, film thickness fluctuations are spontaneously
amplified eventually leading to film rupture and  the formation of
drops/particles with well defined spatial order \cite{Wyart90}. 

Dewetting dynamics leading to particle formation have been studied
in detail in polymer films that are in the liquid state close to room
temperature \cite{reiter92,thiele98,stange97,thiele01}. In comparison,
studying dewetting in metals is challenging due to the high melting
point of most metals. Metal dewetting under conventional long-time
annealing is complicated because various surface processes including
metal-substrate chemical interactions and metal diffusion into the
substrate can influence dewetting \cite{pretorius78,ho93}. On the
other hand, rapid melting techniques such as ion irradiation \cite{Averback00,averback01}
or \emph{ns} pulse laser heating could minimize such undesirable effects
and  permit experiments over practical time scales \cite{Favazza06a}.
Approximately 10 years ago it was shown that \emph{single} \emph{ns}
laser pulse melting of various metals (Au, Cu, Ni) on a \emph{Cr metal
layer} resulted in dewetting patterns with spatial order \cite{bischof96b,Herminghaus98}.
However,  pattern formation and self-assembly of spatially ordered
patterns under pulsed laser dewetting of metal films deposited directly
on SiO$_{\text{2}}$ (a substrate of choice for many opto-electronic
applications) have not been pursued in great detail, especially in
nanoscopic metal films with thickness $h$ $<$ the laser absorption
depth $\alpha_{m}^{-1}$. Furthermore, pattern evolution under the
application of thermally-independent multiple pulses (i.e. the film
undergoes melting during the pulse and resolidification between pulses)
with varying laser parameters to understand the dynamics of metal
dewetting as well as the length scales of the variety of different
patterns have  not been investigated. This is of fundamental importance
since the \emph{unique} thermophysical properties of metallic liquids,
including their high surface tension $\gamma$ of $O(1)\, J/m^{2}$
and large thermal conductivities, distinguish them from the typical
liquid polymers ($\gamma\sim O(10^{-2})\, J/m^{2}$) investigated
in the past.  Therefore, liquid metals could present novel patterning
mechanisms and/or characteristics. Another important question of broader
fundamental interest is related to the applicability of \emph{continuum-level}
conservation equations (for heat, mass and momentum) at the nanoscopic
length scales (\emph{h} < $\alpha_{m}^{-1}\sim11$ nm) used in the
present experiments. Note that the application of  atomistic or mesoscopic
techniques to problems which involve phase change and surface deformation
under fast thermal cycles is not entirely straightforward and  first
principles-based atomistic/mesoscopic simulations of such phenomena
are beyond present capabilities. Hence, the importance of developing
experimentally validated continuum-level models for such processes
cannot be over-emphasized.

Recently we showed that multiple \emph{$ns$} pulsed laser melting
of nanoscopic Co films leads to short- and/or long-range spatial order
\cite{favazza06d,favazza06c,Favazza06b}. We provided preliminary
evidence that dewetting by multiple \emph{ns} pulse laser melting
shows characteristics of a thin film hydrodynamic (TFH) instability,
such as the spinodal process \cite{favazza06d}. This was based on
the observation that the final nanoparticle morphology, quantitatively
described by  the nearest-neighbour (NN) particle spacing ($R$) and
diameter ($D$), is consistent with the prediction of the linear TFH
theory which was modified to incorporate long range inter-molecular
attractive forces of van der Waal's type. However, the quantitative
role of various laser parameters including the energy density $E$
and time (as measured implicitly by the number of pulses $n$) on
pattern evolution was not investigated in detail. In this work we
present a detailed account of pattern formation and dewetting under
\emph{multiple ns} pulsed laser irradiation of Co nanofilms ($h\leq7\, nm$
< $\alpha_{m}^{-1}\sim11\, nm$) on SiO$_{\text{2}}$ in the light
of first-principles thermal modeling and TFH theory (Sec. \ref{sec:Experimental-results-and})
and discuss novel physical mechanisms specific to dewetting of nanolayers
of liquid metals. 

The remainder of the paper is organized as follows. In Sec. \ref{sec:Pulsed-Laser-Processing:}
the experiment and modeling details are presented. In Sec. \ref{sec:Laser-energy-threshold}
we show that by accounting appropriately for the changes in the reflectivity
of the substrate-metal bilayer with respect to \emph{h}, quantitative
agreement can be obtained between experimental observations and the
predictions of continuum-level theory for the \emph{h}-dependence
of the melting energy threshold, pattern length scale and particle
size. This observation corroborates atomistic simulations of heat
transport for simpler fluids under shear flow which, for liquid layers
with several (6-7) atomic layers \cite{khare97}, yielded thermal
profiles consistent with predictions of continuum-level analysis.
For the metal films used in this work, this translates to \emph{h}
< 1 nm.  The thermal model also predicted the development of intrinsic
thermal gradients $\partial T/\partial h$ whose sign is positive
for $h<$ a critical value $h_{c}\approx9-13\, nm$ and negative for
$h>h_{c}$. This would imply that thermocapillary effects (i.e. flow
induced by surface tension gradient along the perturbed interface
which samples an \emph{h}-dependent temperature field) could also
be an important mechanism of pattern formation. For the experimental
thickness regime investigated, the thermal diffusion time scales $\tau_{H}$
$\leq$ pulse time $t_{p}=9\, ns$ and therefore the influence of
$\partial T/\partial h$ was observed to be minimal.  However, since
$\tau_{H}\propto h^{4}$, thermocapillary mechanisms could be important
for thicker films, a hypothesis that needs to be verified by future
experimental studies. In Sec. \ref{sub:Dewetting-pattern-evolution},
we discuss the quantitative behavior of pattern evolution as a function
of  $E$ and  $n$. Extremely fast thermal cycles (heating/cooling
rates of O($10^{10}$K/s)) were predicted by the thermal modeling,
implying that each laser pulse was thermally independent of one another
(repetition rate: 50 Hz). Furthermore, the model predicted liquid
lifetimes $\tau_{L}$ of $2-15\, ns$, while the linear TFH theory
\cite{vrij66} predicted a dewetting time scale $\tau_{D}\geq25\, ns$
implying that multiple pulses would be required to observe the instability,
as confirmed by experiments. We suggest that once dewetting was initiated
by melting of the thin film, subsequent pulses continued to foster
the instability and each laser pulse captured snapshots of its progression.
Experimentally, the dewetting patterns were characterized by short
range order (SRO) at all observed stages of dewetting with the patterns
typically evolving with increasing number of laser pulses $n$ from
discrete holes to bicontinuous or cellular structures and eventually
into nanoparticles. These nanoparticles form by capillary flow rather
than a Rayleigh-like breakup typically observed in polymer liquids.
This behavior can be attributed to the large capillary velocity (ratio
of interfacial tension $\gamma$ to fluid viscosity $\eta$) for the
metal nanolayers resulting from its unique thermophysical parameters,
including a large $\gamma=1.88\, J/m^{2}$, small $\eta=4.45\times10^{-3}Pa-s$
and small pattern length scales due to its large Hamaker constant
$A\sim10^{-18}\, J$. This final nanoparticle state was  robust, characterized
by SRO in NN spacing $R$ and a monomodal size distribution with diameter
$D$ with the observed trends  of $R\sim h^{2}$ and $D\sim h^{5/3}$
respectively. This robustness could be explained by two effects: (i)
the relatively weak temperature dependence of the thermophysical parameters
of Co; and (ii) the minimal particle coarsening due to surface diffusion
since the diffusion time scale is much greater than that of the thermal
processing time $t_{p}$. In Sec. \ref{sec:Role-of-Co} we show that
the pattern ripening rate $L/n$, where $L$ is the characteristic
pattern length, increased linearly with $E$ at the early stages of
patterning which could be explained by the predicted linear increase
in $\tau_{L}$ with $E$.  \emph{Overall, these results show that
pattern formation and dewetting in liquid metal nanolayers under fast
thermal cycles can result in novel mechanisms and, experimental observations
in conjunction with theoretical analyses of thermal cycles and TFH
instabilities offer  quantitative guidelines for  achieving  desired
quenched-in  morphologies with predictable length scales. }

\section{Pulsed Laser Processing: Experiment and Thermal Modelling\label{sec:Pulsed-Laser-Processing:}}

\subsection{Experimental details\label{sec:Experimental-details}}

Cobalt metal films with thickness ranging from $1\leq h\leq7\, nm$
were deposited at the rate of $1\, nm/min$ onto optical quality SiO$_{\text{2}}$/Si
substrates at room temperature. The substrates were commercially obtained
with a $400\, nm$ thick thermally grown oxide layer on polished Si(100)
wafers. Deposition was done under vacuum conditions $(2\times10^{-8}\, Torr)$
by e-beam evaporation from a TECTRA \emph{$e^{-}$-flux} mini electron-beam
evaporator. The wafers were degreased in acetone, methanol, and DI
water prior to film deposition. The film thickness was controlled
through the deposition rate as measured using an in-situ Inficon XTM/2
deposition monitor. The final thickness was established by calibrating
concentration measurements of Co film using energy dispersive X-ray
spectrometry (EDS) in a scanning electron microscope (SEM) to measurements
by Rutherford backscattering \cite{zhang03,zhang05a}. During each
deposition run, up to 8 pieces of the substrate measuring approximately
$5\times5\, mm^{2}$ were placed such that a number of different irradiation
experiments could be performed in vacuum. The thickness variation
between the different pieces in a single run was measured to be less
that 10\%. For every film thickness we also made atomic force measurements
(AFM) of the surface roughness and an upper limit of $0.1\pm.03\, nm$
was established for the average rms roughness over the entire film
thickness range. AFM was performed in ambient conditions using a Molecular
Imaging PicoScan AFM with a tip of spring constant 0.2 N/m. The measured
lateral ($x-y$) resolution was $\sim$ 10 nm and $z$-height resolution
was $\sim$ 0.2 nm. A Hitachi S-4500 field-emission microscope operating
at 15 kV was used for EDS and SEM measurements. The spatial resolution
of this microscope was $\sim1.5\, nm$. 

The films were irradiated in vacuum at normal incidence by a laser
beam of variable beam size controlled by a 500 mm biconvex lens, with
a maximum size of $3\times3\, mm^{2}$. The laser used was a Spectra
Physics Injection seeded Lab-130-50 Nd:YAG laser operating at its
4$^{\textrm{th}}$ harmonic with wavelength $\lambda=266\, nm$ at
$50\, Hz$ repetition rate. Based on the thermal model (Sec. \ref{sub:Dewetting-pattern-evolution}),
the metal film is expected to return to room temperature within approximately
$500\, ns$ and therefore each laser pulse could be viewed as being
thermally independent of one another. The laser beam was plane polarized
with a coherence length of $\sim2\, m$ and had a Gaussian temporal
shape. The pulse width $t_{p}$, as measured by the full-width-half-maxima
of the temporal profile, was $9\, ns$. The spatial profile of the
beam was a circular multimode energy distribution established by the
diffraction output coupler of this laser. The spatial energy distribution
was best described by fitting to a Gaussian shape which gave a 85\%
fit to the centroid of the distribution in each direction with a full
width at half maxima of $4\, mm$. In the studies reported here, all
morphology investigations were reported from an approximately 50 $\times$
100 $\mu m^{2}$ region of the irradiated area within which the laser
energy density variation was $\leq$10\% in each of the two dimensions.

Two types of laser irradiation experiments were performed. The first
was determination of the melt threshold laser energy density, $E_{m}$.
We defined $E_{m}$ as the minimum energy density at which a morphology
change could be detected after the shortest irradiation time of  $10$
laser pulses. We also verified that for $E<E_{m}$ no morphology change
could be observed even after the longest irradiation time of 10500
pulses, thus ruling out contributions from surface diffusion. The
morphology change observed was a typical dewetting pattern observed
in the SEM, as shown in Sec. \ref{sub:Dewetting-pattern-evolution}.
 Such an approach has been successfully used in the past to detect
melt threshold of thin films under laser irradiation \cite{matthias94}.
The laser energy density was controlled by varying the lens-sample
position and/or the laser energy, which was measured using an Ophir
electronics \emph{PE30 A-P} high-power laser detector head with an
ORION power meter. In this manner energy densities ranging from $10-1000\, mJ/cm^{2}$
can be achieved. The second study was of dewetting pattern evolution
as a function of various energy densities $E\geq E_{m}$ and over
a range of irradiation times, as measured by the number of pulses
$n$. The irradiation time typically ranged from $10\leq n\leq10500$.
The pulse number was controlled by a shutter in the beam path. Each
independent irradiation i.e., with a different $E$ and/or $n$ were
performed on different pieces of the substrate and each was characterized
using SEM and/or AFM measurements. While the lower limit on the choice
of $E$ was set by $E_{m}$, the upper limit was set by ensuring that
after the longest irradiation times no substantial evaporation $(\leq15\%)$
had occurred. This was ensured by measuring the concentration of the
metal after irradiation via EDS in the SEM and comparing with the
thickness calibrated values. In this manner detailed information on
pattern formation was obtained as a function of $E$ and $n$ for
each $h$ under conditions not influenced by evaporation.

\subsection{Thermal transport modelling \label{sec:Analytical-Model-of} }

\noindent Given the applicability of continuum-level modeling established
in this work, we point out that the nanoscopic length scales allow
for certain simplifications, which in turn allowed for obtaining insightful
analytical solutions for the film temperature as a function of time.
The simplifications include: (i) $h<\alpha_{m}^{-1}$ and (ii) the
heat loss occurred primarily through conduction through the substrate.
Typically for pulse durations > 1 ns, Fourier heating theory is found
to be applicable and consequently the laser heating may be modeled
by a position-dependent source term in the transient energy equation
\cite{yilbas99}. The heat transport model was developed to simulate
the experiment of heating by a pulsed laser beam incident normal to
the film-substrate bilayer with a spatially uniform energy profile
in the plane ($x-y$) of the bilayer that was infinitely wide in the
$x$ and $y$ directions. For these conditions, the relevant heat
diffusion was normal to the film surface in the $z$-direction where
 $z=-h$ corresponds to the vacuum-film interface and $z=0$ the film-substrate
interface.  The model simulated a 1-D geometry consisting of a thin,
planar and continuous metallic layer on a thermally insulating substrate
with the schematic shown in Fig. \ref{cap:Schematic}(a). The following
assumptions were made in formulating the heat transfer problem:  (1)
The substrate layer was viewed as a semi-infinite medium ($0\leq z<\infty$).
 In the case of SiO$_{\textrm{2}}$ for UV irradiation, the substrate
served strictly as a conducting medium since it was transparent to
UV, and (2) The thermal contact resistance at the film-substrate interface
was neglected which is justified aposteriori as follows. The numerically
predicted thermal flux at the film substrate interface is of order
$\sim10^{10}W/m^{2}$ for films irradiated with laser energy densities
in this work ranging from $\sim50-350\, mJ/cm^{2}$. Typical values
of thermal resistance for thin films range from $10^{-5}$ to $10^{-11}$$Km^{2}/W$
\cite{semmer04}, implying that an addition of thermal impedance would
lead to model predictions for the melting threshold which are significantly
lower that its experimentally measured values. 

\noindent In the above framework, the film temperature  $T^{*}(z,t)$
is described by:\begin{equation}
(\rho C_{eff})_{m}\frac{\partial T*}{\partial t}=k_{m}\frac{\partial^{2}T*}{\partial z^{2}}+S^{*}f(t)\cdot\alpha_{m}\exp(-\alpha_{m}(z+h))\label{eq:filmheateq}\end{equation}
where the subscripts $m$ and $s$ denote the metal and substrate
respectively, $\rho$ is the density, $C_{eff}$ is the effective
heat capacity, $S^{*}$ is the rate of energy flux absorbed from the
laser into the film, $f(t)$ denotes the temporal energy distribution
of the laser source and $\alpha_{m}$ is the absorptivity of the metal
film at $266\, nm$. Specifically,  $f(t)=1$, $1-H(t-t_{p})$ and
$e^{-\frac{(t-t_{p})^{2}}{2\cdot\sigma^{2}}}$ for uniform,  square-shaped
and Gaussian pulses respectively with $H$ denoting the Heaviside
step function. We are interested in the average film temperature especially
since the spatial variations in $T^{*}$ are expected to be small
since $h<\alpha_{m}^{-1}$ and thermal diffusion time scale < $t_{p}$.
An equation for $T\equiv\frac{1}{h}\int_{-h}^{0}T^{*}dz$ can be obtained
by integrating Eq. \ref{eq:filmheateq} with respect to \emph{z} from
$z=-h$ to $z=0$ and dividing throughout by $h$:\begin{equation}
(\rho C_{eff})_{m}\frac{dT}{dt}=-\frac{q_{s}}{h}+\frac{S^{*}f(t)}{h}\left(1-\exp(-\alpha_{m}h)\right)\label{eq:averagedfilmeq}\end{equation}
where $q_{s}(t)\equiv-k_{m}(\partial T^{*}/\partial z)_{0}=-k_{s}(\partial T_{s}/\partial z)_{0}$
is the conductive heat transfer to the substrate where $T_{s}$ denotes
the substrate temperature field. Note that the radiative heat transfer
from the film surface to the vacuum is typically much smaller than
the conductive transport to the substrate. Hence, $(\partial T^{*}/\partial z)_{z=-h}$
is set to 0. Thus, the lumped parameter equation is valid for all
$h$ given that $T$ is interpreted as the average temperature in
the film. Evidently, for  $h<\alpha_{m}^{-1}$, the gradients would
be sufficiently small so that this would be a good approximation to
the uniform film temperature.  Letting $S=\frac{S^{*}(1-\exp(-\alpha_{m}h))}{(\rho C_{eff})_{m}h}$,
the time-dependent temperature in the metallic monolayer can be expressed
as:\begin{equation}
\frac{dT}{dt}=S\cdot f(t)-\frac{q_{s}(t)}{(\rho C_{eff})_{m}h}\label{eq:film_analytic}\end{equation}

\noindent \begin{flushleft}where $S$ is the rate of energy absorbed
from the laser, averaged over the film.  For a Gaussian shaped pulse
we have:\begin{equation}
S^{*}=\left((1-R(h))\cdot\frac{E_{o}}{\sqrt{2\cdot\pi}\cdot\sigma}\right)\label{eq:filmsourceGuass}\end{equation}
\par\end{flushleft}

\noindent and for constant heating or a square pulse\begin{equation}
S^{*}=\left((1-R(h))\cdot\zeta\cdot\frac{E_{o}}{t_{p}}\right)\label{eq:filmsourcesquare}\end{equation}

\noindent where $R$ is the \emph{h}-dependant reflectivity for the
laser at normal incidence. The general expression for the effective
$R$ value for an absorbing film on a transparent substrate has been
derived previously \cite{Heavens55}. Fig. \ref{cap:Schematic}(b)
presents the change in $R$ for Co-SiO$_{\text{2}}$ over the film
thickness range of interest, and could be well represented by a saturating
exponential of type $r_{o}(1-e^{-a_{r}h})$, with $r_{o}=0.44$ and
$a_{r}^{-1}\sim15.5\, nm$. Further, in the above expressions, $E_{o}$
is the laser energy density, $\sigma$ is the standard deviation chosen
so that the length of the Gaussian full-width half-maximum is equal
to the laser pulse width $t_{p}$, $\zeta$ is a renormalization factor
to ensure that the total energy absorbed per pulse is identical for
the square and Gaussian models. The temperature field in the substrate
was described by the variable $\Theta_{s}\equiv T_{s}-T_{0}$, where
$T_{0}=T_{s}(z\rightarrow\infty)$: 

\noindent \begin{equation}
\frac{\partial\Theta_{s}}{\partial t}=a_{s}\frac{\partial^{2}\Theta_{s}}{\partial z^{2}}\label{eq:Laplace_subheat}\end{equation}

\begin{flushleft}where the substrate thermal diffusivity $a_{s}=\frac{k_{s}}{(\rho C_{eff})_{s}}$
with initial/boundary conditions $\Theta(z,\,0)=\Theta(\infty,\, t)=0$
and $-k_{s}\frac{d\Theta}{dz}(0,\, t)=q_{s}(t),$ which couples the
substrate and film temperature fields. By employing the method of
Laplace's transforms, the above set of equations can be solved to
yield: \begin{equation}
T(t)=T_{o}+\frac{2}{\sqrt{\pi}}\cdot\left(\frac{S}{K}\right)\sqrt{t}+\frac{S}{K^{2}}\cdot\left(e^{K\,^{2}t}\cdot\textnormal{erfc}(K\sqrt{t})-1\right)-S\cdot\int_{0}^{t}e^{K^{\,2}(t-u)}\cdot\textnormal{erfc}(K\sqrt{t-u})\cdot H(u-t_{p})\cdot du\label{eq:analytic_solsqpls}\end{equation}
\par\end{flushleft}

\noindent where $K=\frac{\sqrt{(\rho C_{eff}k)_{s}}}{(\rho C_{eff})_{m}h}$
and the integral in Eq. \ref{eq:analytic_solsqpls} may be solved
numerically. Now for the case of the Gaussian model, the solution
for \emph{T}(\emph{t}) is given by: \begin{equation}
T(t)=T_{o}+S\cdot\int_{0}^{\, t}\exp\left(-f^{\,2}\cdot(t-u)^{2}+g\cdot(t-u)+K^{\,2}\cdot u\right)\cdot\,\textnormal{erfc}(K\sqrt{u})\cdot du\label{eq:analytic_sol}\end{equation}

\noindent where the integral in Eq. \ref{eq:analytic_sol} may be
solved numerically using standard  software packages. For both the
Gaussian and square pulse solutions, we used the Riemann midpoint
rule to evaluate the integral with Maple$^{\text{TM}}$  v9.1. We
note that for a uniform pulse ($f(t)=1$), a simpler analytical solution
of the following form can be obtained: \begin{equation}
T(t)=T_{o}+\frac{2}{\sqrt{\pi}}\cdot\left(\frac{S}{K}\right)\sqrt{t}+\frac{S}{K^{2}}\cdot\left(e^{K\,^{2}t}\cdot\textnormal{erfc}(K\sqrt{t})-1\right)\label{eq:ft1}\end{equation}
which for $t>1/K^{2}$ (where $1/K^{2}$ ranges from $4.8\cdot10^{-3}\, ns$
to $5.9\cdot10^{-1}\, ns$ corresponding to Co film thickness of $1$
to $11\, nm$) would predict a temperature rise proportional to $\sqrt{t}$.
While this solution is insightful, it over predicts the temperature
rise within the film. Hence, a more realistic Gaussian temporal distribution
has been employed in the subsequent analysis.

In order to validate the assumptions made in the analytic treatment,
a two-phase transient heat transfer model  was also solved numerically
using the finite element method by using  the commercial software
FEMLab$^{\text{TM}}$ v3.1. Here, the assumption of homogeneous temperature
in the film was relaxed. In order to mimic the semi-infinite substrate
phase, the boundary condition $T_{s}(z,t)\mid_{z=5\cdot L_{th}}=T_{o}$
was imposed where $L_{th}$ is the thermal diffusion length of the
substrate  $\approx\sqrt{t_{p}a_{s}}$. The  simulations were performed
for different spatial and and temporal resolutions to ensure numerical
convergence.   For temperature independent thermophysical parameters,
we observed good agreement between the analytical and numerical results
for \emph{T} within 0.5 K (see Fig. \ref{cap:Comparison-of-analytic-numeric}(a)).
First order phase transition was modeled by using an effective heat
capacity technique based on previous approaches \cite{Voller87,LewisFEM04},
namely, $C_{eff}=C_{p}+L\frac{\partial F}{\partial T}$, where $F(T)=\left\{ \begin{array}{cc}
\frac{T}{2\varepsilon} & \:\;\; for\,\, T_{m}-\varepsilon<T<T_{m}+\varepsilon\\
0 & otherwise\end{array}\right\} $, $L$ is the latent heat of transformation for the film and $T_{m}$
is the melting temperature of the film. The functional form of $F(T)$
was chosen so that $C_{eff}$ models a first order phase change i.e.,
when $\frac{\partial F}{\partial T}$ is integrated from $T_{m}-\varepsilon$
to $T_{m}+\varepsilon$ the result equals unity. Here, $\varepsilon$
is a small temperature parameter whose value was chosen so that no
oscillation in \emph{T} would be observed during the melting and freezing
transitions. A value of $\varepsilon=15\,\textnormal{K}$ was sufficient
to ensure oscillation-free results in the simulations reported here.
 The temperature dependent and independent materials parameters are
tabulated in Table \ref{tab:Materialparameters}. All simulations
predicted the maximum temperature difference between the film-surface
and the film-substrate interface to be $\ll1\textnormal{K}$. This
indicated that the \emph{z}-averaged temperature field obtained from
the analytical model can be interpreted as a nearly uniform film temperature.

\section{results and discussion \label{sec:Experimental-results-and}}

\subsection{Laser energy threshold $E_{m}$ to observe dewetting vs. thickness
 \label{sec:Laser-energy-threshold}}

\subsubsection{Experiment}

The morphology of the films was first investigated as a function of
increasing laser energy density over a practical time scale ranging
from a few laser pulses to a maximum of $10,500$ laser pulses. The
first observation was that below a critical \emph{E} no perceptible
morphology change could be observed even after the longest irradiation.
On the other hand, above this energy, significant morphology changes
could be observed even after the shortest irradiation time ($n=10$).
This sharp threshold behavior implied that a phase-change process,
such as melting, rather than surface diffusion was responsible for
the morphology change. Using this approach, a measurable critical
$E$ was obtained for a range  of film thickness. The experimentally
determined behavior of $E_{m}$ vs. $h$ is plotted in Fig. \ref{cap:Comparison-of-analytic-numeric}(a)
(filled circles).  This observed trend i.e., a significant increase
in $E_{m}$ with decreasing $h$ for the ultrathin films ($h\leq5\, nm$)
is consistent with the prediction of \cite{matthias94} and the underlying
mechanisms that contribute to this behavior are discussed below.

\subsubsection{Modeling}

In order to determine whether the energy $E_{m}$ closely correlated
with the melt threshold, we used the analytical and numerical models
to estimate the melt threshold energy. This was obtained by simulating
the thermal behavior under the pulse for various energies. Fig. \ref{cap:Comparison-of-analytic-numeric}(b)
shows a plot of Eq. \ref{eq:analytic_sol} versus time for various
Co film thicknesses $h$ with a laser energy density value of $E=100\, mJ/cm^{2}$
estimated using the analytic model. The points represent Eq. \ref{eq:analytic_sol}
solved at various times and the lines represent the corresponding
finite element simulations. Predictions of a numerical simulation
incorporating phase change and temperature-dependent material parameters
are shown in Fig. \ref{cap:Comparison-of-analytic-numeric}(c). In
addition, from Eq. \ref{eq:analytic_sol}, the melt threshold energy
$E_{m}$, which is the value of $E$ needed to bring the Co film  to
melting temperature was calculated. The time $t_{m}$ where Eq. \ref{eq:analytic_sol}
reaches its maximum temperature was first determined for the various
film thicknesses. Then, Eq. \ref{eq:analytic_sol} was set equal to
$T_{m}$ at $t=t_{m}$ and solved for the appropriate value of $E$
by successive approximations. Analytical predictions for the melt
threshold energy versus $h$ is shown in Fig. \ref{cap:Comparison-of-analytic-numeric}(a)
as the solid line along with the numerically predicted laser melt
threshold energy density (dashed line). From Fig. \ref{cap:Comparison-of-analytic-numeric}(a),
an increasing $E_{m}$ with decreasing $h$ is obtained for experiment
and modeling and furthermore, the estimated $E_{m}$ is in excellent
agreement with experiment. Based on this result, we interpreted the
experimentally measured $E_{m}$ to be the melt threshold energy and
all dewetting behavior was investigated for energies at or above the
$E_{m}$ for the various $h$. 

Eq. 9 can be used to explain the intriguing thermal behavior. Specifically,
the temperature rise is proportional to $S/K\propto f_{1}(h)\, f_{2}(h)$
where $f_{1}(h)\equiv1-\exp(-\alpha_{m}h)$ and $f_{2}(h)\equiv1-r_{0}+r_{0}\exp(-a_{r}h)$
represent the contributions to the source term (\emph{S} in Eq. \ref{eq:film_analytic})
from finite laser absorption depth of the metal and metal/substrate
bilayer reflectivity respectively. It is easily verified that  $f_{1}$
increases while $f_{2}$ decreases with increasing \emph{h}. However,
since $\alpha_{m}>a_{r}$, the net effect is an increase in $S/K$
with increasing \emph{h}. Consequently, thicker films will experience
faster heating rates. However, note that if one disregards the latter
effect (of reflectivity), the model would over predict the temperature
rise by approximately 6 and 26 percent for \emph{h} = 2 nm and 10
nm respectively. This suggests that appropriately accounting for the
thickness dependence of the reflectivity, overlooked in previous work
\cite{HenleyPRB05}, is important to arrive at   predictions which
are quantitatively consistent with experimental measurements. Moreover,
the sensitivity of \emph{S}/\emph{K} to laser-metal interaction parameters
implies that the thermal behavior and hence the pattern formation
itself could be highly system dependent.  As shown in Fig. \ref{cap:Schematic}(b),
the thickness-dependent reflectivity of the metal-substrate bilayer
 varied from the value for  bulk SiO$_{\text{2}}$ of $R\sim0$ for
$h_{m}=0$ to the value for  bulk Co of $R=0.44$ \cite{OpticConst}
and in the thickness regime of interest in this work, i.e. $h_{m}\leq\alpha_{m}^{-1}$,
the calculated value is significantly smaller than that of bulk Co.

Another important  result obtained from the thermal modeling is the
prediction of an intrinsic thermal gradient $\partial T/\partial h$
that can result in temperature variations along the film whenever
the irradiated film has height fluctuations, i.e. employing a local
approximation, $\partial T/\partial x=(\partial T/\partial h)\,\partial h/\partial x$.
The length scale of the interface height fluctuations (or the interface
roughness) \textasciitilde{} $\sqrt{k_{B}T/\gamma}\approx0.2\, nm$
for a liquid Co film at 1700 K. While such perturbations are randomly
distributed, TFH theory predicts that a wavelength with fastest growing
timescale (or eigenvalue, see Sec. \ref{sub:Quantitative-evidence-for}
below) will be preferentially amplified and become the dominant, observable
length scale. Temperature variations could result in surface tension
gradients along the interface which in turn will induce thermocapillary
flow that can reduce ($\partial T/\partial h$ > 0) or enhance ($\partial T/\partial h$
< 0) the dewetting rate via thermocapillary flows that promote fluid
motion from the crest (trough) to the trough (crest) of the perturbed
interface. Specifically, $\partial T/\partial h$  can be estimated
from Eq. \ref{eq:ft1} and the resulting dependence on $h$ is shown
in Fig. \ref{fig:dToverdh}, where the ordinate $\frac{\partial T}{\partial h}$
is scaled by the numerical factor $(A^{2}/P)10^{-9}$, where $A=\frac{\sqrt{(\rho C_{eff}k)_{s}}}{(\rho C_{eff})_{m}}$
and $P=\frac{\zeta E_{o}}{(\rho C)_{m}t_{p}}$ are material and laser
related parameters. As seen from Fig. \ref{fig:dToverdh}, the rate
of change in temperature, evaluated for different time-to-melt of
the films, is a strong function of film thickness. One can clearly
note that $\partial T(t,h)/\partial h$ becomes negative at a certain
critical thickness $h_{c}$ when the film melts over reasonable time
scales. $h_{c}$ is observed to lie between $\sim9-13\, nm$ corresponding
to melting times between $1-9\, ns$. The time scale over which $\partial T/\partial h$
will be effective is the thermal diffusion time scale $\tau_{H}$
in the Co metal films and this time should be compared to pulse time
$t_{p}$. Clearly if $\tau_{H}/t_{p}\leq1$, then thermal diffusion
would smooth out  $\partial T/\partial h$ effects over the time of
heating and its role on pattern formation would be negligible. As
we show in Sec. \ref{sub:Quantitative-evidence-for} the metal dewetting
is most likely due to a TFH hydrodynamic instability similar to spinodal
dewetting. According to this this model, the dewetting length length
scale $\Lambda$ varies as $h^{2}$ and thus the relevant thermal
diffusion time is $\tau_{H}\propto\Lambda^{2}/D_{th}^{Co}\propto h^{4}$,
where $D_{th}^{Co}=2.7\times10^{-5}m^{2}/s$ is the thermal diffusivity
of Co. Using the quantitative expression for $\Lambda$ obtained in
Sec. \ref{sub:Quantitative-evidence-for} we determined that up to
thickness of $h\sim6\, nm$ $\tau_{H}\leq t_{p}$ and therefore the
intrinsic thermocapillary effects will only have a minimal influence
of pattern formation for the thickness regime investigated here $1\leq h\leq7\, nm$.
On the other hand, since $\tau_{H}\propto h^{2}$ significant modifications
to pattern formation for thick films is likely, including changes
to patterning length and time scales. \emph{}However, a detailed investigation
for thickness above $7\, nm$ is beyond the scope of the present work.

\subsection{Dewetting pattern evolution with $n$ and $E$ \label{sub:Dewetting-pattern-evolution}}

A detailed investigation of the dewetting morphology was performed
for various laser energies as a function of the laser irradiation
time $n$. For films in the range of $3\leq h\leq7\, nm$ the patterns
typically consisted of discrete holes at the early stages of irradiation,
followed by cellular patterns at later stages and eventually nanoparticles
which remained stable to irradiation. On the other hand, for the films
with $h<3\, nm$, the morphology consisted of discrete holes followed
by a bi-continuous structure with a final state again characterized
by nanoparticles.  It is widely accepted that the dewetting morphology
can arise from three mechanisms \cite{degennes03}. (i) Homogeneous
nucleation and growth in which holes appear randomly in location and
time on the surface.  Therefore, no characteristic length is present
in this type of dewetting \cite{stange97}. (ii) Heterogeneous nucleation
and growth due to defects, impurities or other experimentally imposed
heterogeneities. Here, a characteristic length scale could appear
at the early stages of dewetting due to ordered nucleation sites.
For instance, in ion-irradiation induced dewetting, the average molten
zone of an ion imposes a characteristic length scale in dewetting
\cite{averback01}. (iii)  TFH instabilities  such as the one associated
with the dewetting of spinodally unstable systems. The resulting patterns
 were characterized by a well-defined length scale in the hole spacing
and/or size \cite{thiele98}. Since we have ruled out spatially ordered
heterogeneities on the surface and in the film microstructure, the
results presented below were analyzed to distinguish between dewetting
by homogeneous nucleation and  TFH  instabilities that establish a
characteristic length scale. In order to do this, the primary characterization
was evaluation of the patterning length scales obtained from the power
spectrum (PS) evaluated by performing fast fourier transforms (FFT)
of the morphology of the dewetting patterns.

\subsubsection{Dewetting morphology for $h<3\, nm$\label{sub:Dewetting-morphology-for}}

Fig. \ref{cap:2.4nmpatterns}(a) to (d) capture the the morphology
of a 2 nm thick Co film as a function of increasing number of laser
pulses at energy  $200\, mJ/cm^{2}$. Discrete holes were visible
after the shortest time (Fig. \ref{cap:2.4nmpatterns}(a)) with the
patterns changing to a bi-continuous structure (Fig. \ref{cap:2.4nmpatterns}c
and d) and finally into nanoparticles (Fig. \ref{cap:2.4nmpatterns}(d).
A comparison of the density of features in Fig. \ref{cap:2.4nmpatterns}(c)
and (d) indicates that the nanoparticles form by change of spaghetti-like
features into a single particle rather than break-up of the spaghetti-like
features into multiple particles, such as in a Rayleigh-like process.
As shown in Fig. \ref{fig:Monomodal-size-distribution}(a), the nanoparticles
had a monomodal size distribution. The spatial behavior of the  patterns
could be understood from the PS, as shown in Fig. \ref{cap:2.4nmpatterns}(a)
to (d). During all observed stages of the pattern evolution an annular
shaped PS was visible, suggesting that a band of characteristic spatial
frequencies was present in the patterns. Qualitatively, the radius
of the annular region changed with increasing irradiation time reflecting
the change in morphology of the dewetting patterns. Fig. \ref{cap:2.4nmPatternvsE}(a)
to (d) show the pattern morphology after 100 pulses as a function
of various energies for the 2 nm film. The general characteristics
of the morphology were similar to those observed as a function of
$n$, as shown in Fig. \ref{cap:2.4nmpatterns} and the primary role
of increasing energy appears to be an enhanced rate of achieving the
nanoparticle state.

\subsubsection{Dewetting morphology for $3\leq h\leq7\, nm$\label{sub:Dewetting-morphology-for}}

For the thicker films the intermediate stages of dewetting differed
considerably in their morphology from Fig. \ref{cap:2.4nmpatterns}
but still retained a characteristic spatial frequency. Fig. \ref{cap:3.5nmpatterns}(a)
to (d) capture the morphology of a $4.4\, nm$ thick Co film as a
function of increasing number of laser pulses at energy $93\, mJ/cm^{2}$
. Discrete holes were visible after the shortest time (Fig. \ref{cap:3.5nmpatterns}(a))
with the patterns changing to a cellular structure (Fig. \ref{cap:3.5nmpatterns}(c))
as the number of holes increases. Continued irradiation results in
the metal being pushed to the edge of the holes and the patterns consisted
of large polygonal structures with evidence for particle formation
occurring preferentially at the vertices. This is more evident in
Fig. \ref{cap:3.5nmpatterns}(c). After long irradiation times, a
stable nanostructure morphology was observed (Fig. \ref{cap:3.5nmpatterns}(d)).
This behavior is consistent with particle formation via capillary
draining of the liquid bridges into the vertices  (Sec. \ref{sub:Quantitative-evidence-for}). 

Analysis of the PS for each pattern again showed an annular structure
indicating spatial order in the dewetting patterns, as shown in Fig.
\ref{cap:3.5nmpatterns}(a) to (d). The radial distribution function
$g(k)$ for each of these PS and the curve fit used to obtain $L$
are shown in Fig. \ref{cap:3.5nmLvsnPS}(a). The cumulative behavior
of $L$ vs. $n$ for this film is shown in Fig. \ref{cap:3.5nmLvsnPS}(b).
As in the case of the $2\, nm$ film, $L$ increases as the holes
appear and merge to form closed ring-like polygons. In this scenario,
$L$ represents the average diameter of the holes or polygons. However,
unlike the case of $2\, nm$ film, the appearance of the nanoparticles
decreases the characteristic length scale. This can be understood
by noting that  the final length scale represents the average NN spacing
of the nanoparticles which formed preferentially at the vertices of
the polygons. Since the average vertex spacing is smaller than the
average diameter of the polygons therefore $L$ decreased as polygons
changed into nanoparticles. The nanoparticles once again had a monomodal
size distribution, as shown in Fig. \ref{fig:Monomodal-size-distribution}(b).
We also investigated the role of laser energy on the pattern formation.
Similar to the case of the $2\, nm$ film, we observed that the primary
effect of increasing energy is an enhanced rate of achieving the nanoparticle
state. This effect is better represented in Fig. \ref{cap:3.5nmLvsnPS}(b)
where the characteristic $L$ determined from the $g(k)$ is plotted
as a function of $n$ for various laser energies $E$. As the laser
energy increases, the observed value of $L$ is larger suggesting
that the pattern ripening rate increased. However,  $R$ which represented
the nanoparticle morphology after long irradiation times was independent
of the laser $E$ (Sec. \ref{sec:Role-of-Co}). This result once again
confirmed that the final nanoparticle state is  robust to the laser
parameters.

\subsubsection{Quantitative evidence for TFH instability \label{sub:Quantitative-evidence-for}}

From the above results we suggest that the pulsed laser melting initiates
the nonlinear dewetting TFH instability  and subsequent pulses evolve
this instability to a well-defined final state\emph{.} Further, each
observed state is characterized by a narrow band of spatial frequencies
indicating spatial order over the entire dewetting process. Since
we ruled out heterogeneous nucleation from ordered surface defects,
the above results point strongly to dewetting by an instability. A
more quantitative evidence for this hypothesis can be pursued on the
basis of predictions of the linear TFH theory according to which the
characteristic time scales and length scales of dewetting are well-defined
functions of $h$. From our previous results, the most robust measure
of the dewetting length scale appears to be the final nanoparticle
state \cite{favazza06d}.  Therefore, we  studied the characteristics
of this final nanoparticle state including, $R,$ D, contact angle
function $f(\theta)$ and the particle areal density $N$ as a function
of film thickness $h$ to determine pattern formation via the TFH
instability. 

\begin{enumerate}
\item $R$ vs $h$

According to the linear TFH instability, the characteristic spinodal
length $\Lambda$ as a function of film thickness can be expressed
as \cite{vrij66,Wyart90,sharma93,favazza06d}: \begin{equation}
\Lambda(h)=\sqrt{\frac{16\pi^{3}\gamma}{A}}h^{2}=Bh^{2}\label{eq:Spinodallength}\end{equation}

where $\gamma$ is the surface tension of the metal and $A$ is the
Hamaker constant. Since we interpret the final nanoparticle morphology
as resulting from the dewetting instability, we assigned the observed
NN spacing $R$ to be proportional to $\Lambda$  as $R=a\Lambda$,
where $a$ is the proportionality factor. This is reasonable given
that the average nanoparticle spacing is related to the average size
and/or spacing between the ordered features arising from the dewetting
process. We determined the behavior of $R$ as a function of film
thickness, as shown Fig. \ref{fig: Spinodal-length-scale}(a). The
power law fit to the spacing $R$ yielded a behavior expressed by
$R(nm)=25.7h^{1.98\pm0.3}$ with the film height expressed in \emph{nm}.
The NN spacings varied from $\sim40$ to $1000\, nm$ for the films
studied. The exponent of the film thickness of $1.98\pm0.3$ is in
excellent agreement to the theoretical value of $2$. From this experimentally
established pre-exponential factor of $25.7\, nm^{-1}$ the Hamaker
constant was estimated to be $A=1.41\times10^{-18}J$ which is of
the right order of magnitude \cite{israelachvili92}. 

\item Dewetting instability  time scale $\tau_{D}$

In the multiple pulse laser melting dewetting scenario, the morphology
change occurs over multiple cycles of phase change, i.e. melting and
resolidification. Clearly, once the instability is initiated, in order
for it to be fostered by the multiple pulses, the time scale of dewetting
$\tau_{D}$ must be larger than the lifetime of the liquid $\tau_{L}$
in each pulse. From the linear TFH theory, $\tau_{D}$ can be expressed
as: \begin{equation}
\tau_{D}(h)=\frac{96\pi^{3}\gamma\eta}{A{}^{2}}h^{5}\label{eq:tauD1}\end{equation}

where $\eta=4.46\times10^{-3}\, Pa-s$ is the viscosity of liquid
Co at its melting point. Using the experimentally determined value
of \emph{A} we find that $\tau_{D}\sim25\, ns$ for a $1\, nm$ thick
film and increases as $h^{5}$. From our modeling, we determined that
at the melt threshold of $\sim338\, mJ/cm^{2}$ for the 1 nm film,
the liquid lifetime is $\sim4\, ns$. The typical liquid lifetime
was obtained from thermal profiles of the type shown in Fig. \ref{cap:Comparison-of-analytic-numeric}(b)
and (c). The liquid lifetime for the analytic result (Fig. \ref{cap:Comparison-of-analytic-numeric}(b))
was defined by placing a line at $T_{m}=1768$ $K$, the Co melting
temperature, and noting the time  between both intercepts. For the
numerical model (Fig. \ref{cap:Comparison-of-analytic-numeric}(c)),
the liquid life time is defined as the time needed for the system
to progress from the onset of melting to the completion of freezing.
From this analysis, the liquid lifetime for the range of films studied
here varied between $2-15\, ns$, as determined from the thermal model,
and clearly is smaller than the  dewetting time estimated from Eq.
\ref{eq:tauD1}. 

We also verified from the thermal modeling that the heating and cooling
rates were extremely rapid and since the time between pulses was large
($20\, ms$), each pulse was thermally independent of the other. To
extract the rates, first, plots of the type show in Fig. \ref{cap:Comparison-of-analytic-numeric}(b)
were generated for Co film thickness of 1, 2, 3, 5, and 11 nm with
laser energy density values chosen so that the peak temperature would
be $\sim2000\pm100\, K$. The heating rate was estimated from the
time $\Delta t$ to reach maximum temperature $T_{m}$ as $(T_{m}-T_{0})/\Delta t$
and the model predicted typical rate of temperature rise of the bilayer
system to be $\sim150\, K/ns$  for the metal-film thicknesses of
interest. For the cooling rates, and a second order exponential fit
to the cooling portion (e.g. Fig. \ref{fig:Heating-cooling}(a)) and
a typical time constant of $\sim130\,\, ns$ was estimated for the
temperature to fall by $1/e$ of its peak value. In addition, within
approximately $500\, ns$ the bilayer system reached $98\%$ of its
final value giving a time averaged cooling rate of $\sim$4~K/ns.
Clearly, each laser pulse could be treated independently of one another.
These time scale estimates support a dewetting scenario in which the
instability is initiated by a number of pulses and then dewetting
progresses by the subsequent pulses to the final stable nanoparticle
state.

\item $D$ and $f(\theta)$ vs $h$

An analytical relationship between the diameter of a particle $D$
and the NN spacing $R$ can be derived based on volume conservation
\cite{reiter92}. Assuming that a circular section of the film with
a volume approximated as a cylinder of diameter $R$ and thickness
$h$ contributes to one spherical nanoparticle of diameter $D$, then
volume conservation suggests that: \begin{equation}
\frac{\pi}{4}R^{2}h=\frac{4\pi}{3}f(\theta)\frac{D^{3}}{8}\label{eq:SpinodDiamet1}\end{equation}

where $f(\theta)$ is the geometric factor based on the particle contact
angle $\theta$ that determines the fraction of the sphere that lies
above the substrate ( for the case of $\theta=180^{o}$, $f(\theta)$=$1$).
Using the characteristic length from Eq. \ref{eq:Spinodallength}
in Eq. \ref{eq:SpinodDiamet1} we get: \begin{equation}
D=(\frac{24\pi^{3}\gamma}{Af(\theta)})^{1/3}h^{5/3}=Ch^{5/3}\label{eq:DSpinDiameter2}\end{equation}

In Fig. \ref{fig: Spinodal-length-scale}(b) the average particle
diameter $D$ is plotted as a function of $h$. $D$ varied from $\sim30$
to $250\, nm$ for the thickness range studied here. The power-law
fit to our experimental data gives $D(nm)=14.8h^{1.6\pm0.3}$ with
the film height expressed in nm. The exponent of $1.6\pm0.3$ is in
excellent agreement with the theoretical estimate of $5/3$. By using
the experimentally determined pre-exponent of $14.1\, nm^{-2/3}$
in Eq. \ref{eq:DSpinDiameter2}, we determined the geometric factor
to be $f(\theta)=0.361$ giving an effective contact angle for the
nanoparticles of $74^{o}$. Given the simplicity of the model relating
the instability length scale to the particle diameter and the uncertainties
in the values of $R$ and $D$ of approximately 30\%, this experimentally
extracted value of the contact angle is in reasonable agreement to
the previously reported value of $100\pm26^{o}$ measured by atomic
force and scanning electron microscopy \cite{favazza06c}. 

\item Areal density $N$ vs $h$

The areal density \emph{N} of particles can be defined as: \begin{equation}
N=1/(\frac{\pi}{4}\Lambda^{2})\label{eq:ArealDensity1}\end{equation}

which using Eq. \ref{eq:Spinodallength} results in a thickness dependence
given by:\begin{equation}
N(h)=\frac{A}{4\pi^{4}\gamma}h^{-4}\label{eq:Arealdensity2}\end{equation}

Plotted in Fig. \ref{fig: Spinodal-length-scale}(c) is the variation
in the observed particle density with $h$. A power law fit gives
$N=1.5\times10^{-3}h^{-3.8\pm0.8}$ $\#/nm^{2}$ with the film height
expressed in nm. The exponent of $-3.8\pm0.8$ is in fair agreement
with the expected value of $4$ (with the uncertainty primarily arising
from propagating errors in the various quantities) while the pre-exponent
is consistent with the value obtained by substituting the experimentally
determined $A$ and $\gamma$ in Eq. \ref{eq:Arealdensity2}. 

\item Effect of large capillary velocity on pattern formation

From the quantitative results above, the classical TFH dewetting instability
appears to be a suitable explanation of the pattern formation in these
nanoscopic Co films in the regime of weak thermocapillary effects.
At this stage it is also important to emphasize the difference between
the dewetting patterning behavior observed in the metal films reported
here to those of typical polymer films \cite{reiter92}. Previous
studies of the break-up of cellular structures made from typical polymeric
liquids show that the sides of the polygons break into a chain of
droplets by the Rayleigh instability and as a consequence the particle
diameter will follow the trend $D\sim h^{3/2}$ \cite{reiter92}.
However, for the metal films investigated here, the final break-up
of the cellular regions most likely occurs via capillary flow of the
liquid into the vertices of the polygon. We arrived at this conclusion
based on the close correlation in the spatial density of the vertices
and particles (Fig. \ref{cap:3.5nmpatterns}(c) and (d)) or the spaghetti-like
features and particles in Fig. \ref{cap:2.4nmpatterns}(c) and (d).
We attribute this to the extremely large capillary velocities \cite{egger99,dirk04}
of the liquid metals given by the quantity $v_{c}=\gamma/\eta$, which
for Co is $\sim422\, m/s$. This implies that capillary draining from
the liquid bridges connecting the vertices (reservoirs) which are
separated by a distance of 1 $\mu m$ (which is smaller than the largest
NN spacing reported in this work) is \textasciitilde{} 2.5 ns. This
is smaller than the (Rayleigh) ripening time ($\tau_{Ray}\sim2\pi(\rho r^{3}/c\gamma)^{1/2}$),
where $\rho$ is the liquid density, r is the cylinder radius and
c is a factor of \emph{O(0.1)} determined by contact angle and boundary
conditions, $ $for surface perturbations of the liquid bridges, which
typically ranged between several to > 10 ns for the bridge sizes observed
here \cite{amarouchene01}. In this respect, the final stages of breakup
is different from that typically observed in common polymer films
investigated, such as polystyrene, in which the cellular rims break-up
via a Rayleigh-like instability \cite{reiter92}. Moreover, we point
out that the Ohnsorge number, $Oh\equiv\frac{\eta}{\sqrt{\rho\gamma D}}$,
that represents the ratio of viscous to capillary forces is typically
<\,{}< 1 for typical O(10 nm) values of the liquid
bridge diameter, \emph{D}. This further underscores the dominance
of strong capillary effects in the breakup of polygonal structures.
For more viscous liquids (e.g. polymers) with typically smaller interfacial
tension as compared to metals, the relative importance of viscous
and capillary forces can change by orders of magnitude, leading to
different mechanisms of breakup, e.g. via Rayleigh-like process.  Another
factor that can influence the particle size distribution is coarsening
due to surface diffusion. %
However, as seen from Fig. \ref{cap:3.5nmLvsnPS}(b) changing the
laser energy, which changes liquid temperature and lifetime (Sec.
\ref{sec:Role-of-Co}), did not change the final spacing. This was
attributed to the fast thermal processing that minimizes particle
coarsening. 

\end{enumerate}

\subsection{Role of Co liquid temperature and lifetime on dewetting \label{sec:Role-of-Co}}

From the experimental results of dewetting presented in the previous
section, the length scales $L$ during the early stages of dewetting
increased with increasing laser energy following similar irradiation
times (Fig. \ref{cap:3.5nmLvsnPS}). From our thermal modeling, increasing
laser energy increased the Co liquid temperature and the liquid lifetime.
Further, the length scale of dewetting, $\Lambda$ and the dewetting
timescale $\tau_{D}$ have temperature-dependent materials parameters
(i.e. $\gamma$ and $\eta$). Therefore the observations of Fig. \ref{cap:3.5nmLvsnPS}
can arise from three factors: (i) from a change in the dewetting length
scale $\Lambda$, (ii) from an increase in liquid lifetime $\tau_{L}$
per pulse, and (iii) from a change in the time scale for ripening
$\tau_{R}$ of the dewetting structures. However, Fig. \ref{cap:3.5nmLvsnPS}
also indicates that changing the laser energy had no influence on
the final nanoparticle characteristic, implying that no permanent
change to $\Lambda$ must occur with change in liquid $T_{L}$. In
this section we show that the change in $\Lambda$ for Co metal is
experimentally indistinguishable over significant T changes (of up
to 500 K above $T_{m}$). From theoretical arguments we found that
the increase in $L$ with increasing $E$ following irradiation by
a small number of pulses $n$ (i.e. early stages) arises  primarily
due to variations in $\tau_{L}$  and its linear behavior with $E$.
This result was also consistent with  experimental observations based
on the measurement of $L$ vs $n$ for various energies and thickness.

\subsubsection{Role of temperature on $\Lambda$}

The T-dependence of the dewetting length $\Lambda$, which comes from
the surface tension $\gamma(T)$ and the Hamaker constant $A(T)$,
can be expressed as:\begin{equation}
\Lambda(h,T)=\sqrt{\frac{16\pi^{3}\gamma(T)}{A(T)}}h^{2}\label{eq:Lambda}\end{equation}

The variation in temperature can be expressed as: \begin{equation}
\frac{d\Lambda}{dT}=\sqrt{16\pi^{3}}\left[\frac{1}{2A^{1/2}}\gamma^{-1/2}\frac{\partial\gamma}{\partial T}-\frac{\gamma^{1/2}}{2A^{3/2}}\frac{\partial A}{\partial T}\right]h^{2}\label{eq:dLambdaoverdT}\end{equation}

Eq. \ref{eq:dLambdaoverdT} can be simplified for metals on SiO$_{\textrm{2}}$
system by noting that the Hamaker constant for metals is virtually
independent of temperature \cite{french00} and so $\frac{\partial A}{\partial T}\sim0$
resulting in:\begin{equation}
\frac{d\Lambda}{dT}=\sqrt{16\pi^{3}}\left[\frac{1}{2A^{1/2}}\gamma^{-1/2}\frac{\partial\gamma}{\partial T}\right]h^{2}\label{eq:dlambdaoverdT2}\end{equation}

For a temperature rise of $\Delta T$ over the melting point, the
change in length can be expressed as: \begin{equation}
\Delta\Lambda(\Delta T)=\frac{\Lambda(T_{m})}{2\gamma(T_{m})}\frac{d\gamma}{dT}\Delta T=-2.66\times10^{-4}\Lambda(T_{m})\Delta T\label{eq:deltaL}\end{equation}

where $\Lambda(T_{m})$ is the length at the melt temperature and
$\gamma(T_{m})$ and $d\gamma/dT$ for Co are taken from Table \ref{tab:Materialparameters}.
From this we see that $\Lambda$ decreases as $T_{L}$ increases,
but this decrease is small and for temperature rises of up to $\Delta T=500\, K$
is only $\sim-13\%$. This change is smaller than the spread in the
experimental values of $R$ of $\pm20\%$ as measured from the $g(k)$.
Therefore, for conditions under which the Co liquid T rise is $\sim500\, K$
above $T_{m}$ the change in length scale will be experimentally indistinguishable.

\subsubsection{Role of T on ripening rate $L/n$}

According to the linear theory TFH theory,  the pattern ripening time
$\tau_{R}$ is expected to be proportional to the dewetting time $\tau_{D}$
\cite{vrij66} and so we can express the \emph{T}-dependent ripening
time as: 

\begin{equation}
\tau_{R}(h,T)=\omega\tau_{D}=\omega\frac{96\pi^{3}\gamma(T)\eta(T)}{A(T)^{2}}h^{5}\label{eq:tauD}\end{equation}

From this we can also define the  ripening rate as $\sigma\propto\frac{\Lambda}{\tau_{R}}$,
which can be evaluated using Eq. \ref{eq:Lambda} and \ref{eq:tauD}
as:\begin{equation}
\sigma(h,T)=\kappa\frac{\Lambda(h,T)}{\tau_{R}(h,T)}=M\gamma(T)^{-1/2}\eta(T)^{-1}A^{1/2}h^{-3}\label{eq:sigma}\end{equation}

where $M=\frac{\kappa}{\omega}\frac{\sqrt{16\pi^{3}}}{96\pi^{3}}$
is a proportionality constant with unknown numerical factors of $\kappa$
and $\omega$. Using these expressions, the experimentally observed
dewetting length L for a film of thickness $h$ at a given laser energy
$E>E_{m}$ following irradiation by n pulses can be expressed as:
\begin{equation}
L(h,E,n)=\sigma(h,E)\tau_{L}(E,h)n\label{eq:L}\end{equation}
where $\tau_{L}$, the lifetime of the liquid, is a function of the
laser energy and film thickness. In order to evaluate the expected
behavior, each of the temperature (and laser E) dependent terms on
the rhs can be evaluated: 

\begin{enumerate}
\item Behavior of ripening rate $\sigma$(\emph{h,T}) for Co:

The temperature dependence of  $\sigma$ can be expressed analytically
as: \begin{equation}
\frac{d\sigma(T,h)}{dT}=-B\frac{A^{1/2}}{\gamma^{1/2}\eta}h^{-3}\left[\frac{1}{2}\frac{\gamma_{T}}{\gamma}+\frac{\eta_{T}}{\eta}\right]\label{eq:delSigma1}\end{equation}

where, $\gamma_{T}=\frac{\partial\gamma}{dT}$ and $\eta_{T}=\frac{\partial\eta}{\partial T}$
and the temperature dependence of the Hamaker constant has been neglected.
The behavior of $\sigma(T)$ with temperature can be evaluated for
Co metal by using the materials parameters from Table \ref{tab:Materialparameters}.
Fig. \ref{cap:Role of Temperature}(a) shows that Eq. \ref{eq:delSigma1}
is well described by a linear behavior which increases with increasing
temperature of the form: \begin{equation}
\sigma(T,h)=\beta A^{1/2}h^{-3}[0.29T-350]=\sigma_{T}(h)T-\sigma_{o}(h)\label{eq:sigmaTNumer}\end{equation}

\item Dependence on liquid \emph{T} and lifetime $\tau_{L}$ on $E$: Thermal
modeling

\noindent From thermal profiles of the type shown in Fig. \ref{cap:Comparison-of-analytic-numeric}(a),
the peak temperature and liquid lifetime were calculated as functions
of $h$ and $E$. Fig. \ref{cap:Comparison-of-analytic-numeric}(b)
shows thermal profiles of the film temperature for various film thickness
for a laser energy density value of $E=125\, mJ/cm^{2}$. The liquid
lifetime for the analytic model was defined by placing a line at $T_{m}=1768$
$K$, the Co melting temperature, and noting the time in between both
intercepts. For the numerical model, the liquid lifetime is defined
as the time needed for the system to progress from the onset of melting
to the completion of freezing. From Fig. \ref{cap:Comparison-of-analytic-numeric}(b)
we see that introducing  phase change into the model causes an appreciable
 change from the predicted temporal profile by Eq. \ref{eq:analytic_sol}.
The portions of the figure where the temperature plateaus represent
the film melting and freezing. Fig. \ref{fig:Heating-cooling}(b)
presents the liquid lifetime and peak temperature versus laser energy
density trends using the numerical and analytic models. From the figure,
the numerical (open symbols) predictions for liquid lifetimes are
about 20\% lower then  those predicted by the analytic model (closed
symbols) for films greater than 1 nm.  Further from Fig. \ref{fig:Heating-cooling}(c),
the peak temperatures predicted by the numerical simulation are lower
by $\sim200\, K$ than those predicted by the analytic model. Both
these differences arise from the fact that the phase change in the
numerical model consumes more energy and takes more time. Fig. \ref{fig:Heating-cooling}(b)
and (c) present the liquid lifetime and peak temperature as a function
of laser energy density above the melt threshold for various film
thicknesses. From thermal cycles of the type shown in Fig. \ref{cap:Comparison-of-analytic-numeric},
a linear relationship was observed between the liquid life-time and
peak temperature for small changes in laser energy density above $E_{m}$
and can be expressed as:

\end{enumerate}
\begin{equation}
\tau_{L}(E,h)=A(h)E-B(h)\label{eq:TauLiq1}\end{equation}
\begin{equation}
T_{L}(E,h)=C(h)E+D(h)\label{eq:Tliq1}\end{equation}

\noindent where the slopes $A$ and C and intercepts $B$ and D are
thickness dependent positive quantities. Their typical values ranged
between: $0.07\leq A\leq0.3\, ns-cm^{2}/mJ$; $-22\leq B\leq-17\, ns$
and $2.3\leq C\leq14.5\, K-cm^{2}/mJ$; and $D\sim301\, K$ for thickness
between $1\leq h\leq11\, nm$ with $<5\%$ difference between the
numerical and analytical estimates. In this laser energy density regime,
while the temperature of the Co film is above the melting temperature,
there is minimal evaporation over the relevant time scales.

Using Eq. \ref{eq:TauLiq1} in Eq. \ref{eq:L} and Eq. \ref{eq:Tliq1}
in Eq. \ref{eq:sigmaTNumer} we can express the rate of change of
dewetting length L with $n$ as:\begin{equation}
\frac{L(E,h)}{n}=pE^{2}+qE+r\label{eq:L(E,h)}\end{equation}
where $p=\sigma_{T}CA$; $q=\sigma_{T}(DA-BC)-\sigma_{o}A$; and $r=-B(\sigma_{T}D-\sigma_{o})$
are the resulting temperature and thickness coefficients. Equation
\ref{eq:L(E,h)} can be further simplified by analysing the magnitudes
of the various coefficients using the range of values for the coefficients
$A$ to $D$ which showed that for films in the range of 1 to 11 nm
at the melting point of Co, the coefficient $p<<q$ and the quadratic
term in Eq. \ref{eq:L(E,h)} can be neglected to give:\begin{equation}
\frac{L(E,h)}{n}=qE+r\label{eq:L(E,h)2}\end{equation}
Therefore the experimentally observed trend in the ripening rate with
laser energy should be linear. As shown in Fig. \ref{cap:Role of Temperature}(b),
the L/n for for various film thickness can be described by linear
fits (dashed lines). The physical significance of this linear behavior
is the weak temperature dependence of the material parameters of Co.
Therefore the primary contribution to the ripening rate due to increasing
energy is the increasing lifetime of the liquid, which as shown in
Fig. \ref{fig:Heating-cooling}(b), varies linearly with E. This result
also implied that understanding the thermal behavior could result
in predictive models describing the quenched-in dewetting morphologies
and length-scales when such nanoscopic films are subjected to multiple
instances of phase change.

\section{Summary and conclusions \label{sec:Summary-and-conclusions}}

We have investigated the dewetting behavior of nanoscopic Co metal
films with thickness $h$ $<\alpha_{m}^{-1}$ on  SiO$_{\textrm{2}}$
substrates following multiple ns pulse laser irradiation using $266\, nm$
UV light through experiments,  first-principles thermal transport
modeling and analysis based on TFH theory.  We determined that by
accounting appropriately for the changes with respect to \emph{h}
in the reflectivity of the substrate-metal bilayer, quantitative agreement
with experiments can be obtained for the \emph{h}-dependence of the
melting energy threshold predicted by \emph{continuum-level} analysis.
This observation corroborates atomistic simulations of heat transport
for simpler systems under shear flow where the results were found
to agree with the predictions of continuum-level analysis for liquid
layers with several  atomic layers. For the metal films used in this
work, this translates to \emph{h} < 1 nm. A novel effect, the development
of intrinsic thermal gradients $\partial T/\partial h$ whose sign
is positive for $h<$ a critical value $h_{c}\approx9-13\, nm$ and
negative for $h>h_{c}$, was predicted by the thermal model. This
effect arises due to the strong film thickness- and time-dependence
of temperature for liquid metal nanolayers and fast thermal cycles.
This intrinsic effect can lead to thermocapillary effects that could
modify pattern formation. However, for the experimental thickness
regime investigated, the thermal diffusion time scales $\tau_{H}\leq$
 pulse time $t_{p}=9\, ns$ and therefore the influence of $\partial T/\partial h$
is minimal. However since $\tau_{H}\propto h^{4}$, thermocapillary
mechanisms could be important for thicker films, a hypothesis that
needs to be verified by future experimental studies. Because of the
extremely fast thermal cycles (heating/cooling rates of O($10^{10}$K/s))
and short liquid lifetimes $\tau_{L}$ of $2-15\, ns$, multiple pulses
were required to observe the thin film instability with characteristics
dewetting times $\tau_{D}\geq25\, ns$. The dewetting patterns showed
short range order and typically evolved with increasing number of
laser pulses from discrete holes to bicontinuous or cellular structures
and eventually into nanoparticles. The large capillary velocity and
short patterning length scales in the metal layer resulted in nanoparticle
formation by capillary flow rather than via a Rayleigh-like breakup
typically observed in polymer liquids. This behavior can be attributed
to the unique thermophysical properties of the metals, including the
large $\gamma$, small $\eta$ and large Hamaker constant $A$ that
result in large capillary velocities and small patterning length scales.
This final nanoparticle state was extremely robust with nearest neighbour
spacing $R\sim h^{2}$ and a monomodal size distribution with diameter
$D\sim h^{5/3}$ respectively. This robustness was due to the relatively
weak temperature dependence of the thermophysical parameters of Co
and small particle ripening rates due to the fast thermal processing.
The experimental ripening rate $L/n$ increased linearly with laser
energy density $E$ at the early stages of patterning and due to the
linear increase in $\tau_{L}$ liquid lifetime with $E$ predicted
by the thermal modeling implying that accurate predictions and control
of the quenched-in dewetting morphologies and length scales can be
achieved in these nanoscopic systems. These results show that nanoscopic
metal film dewetting following fast thermal processing result in novel
quenched-in dewetting morphologies and length scales which can be
accurately predicted by first-principles modeling.  

RK acknowledges support by the National Science Foundation through
a CAREER grant  DMI-0449258. JT acknowledges valuable conversations
with Dr. Massoud Javidi in regards to the phase-change implementation
in the numerical simulations.  RS and DT acknowledge NSF grant  CTS
0335348. \textbf{\large }{\large \par}

\pagebreak

\section*{Table Captions}

\begin{itemize}
\item Table \ref{tab:Materialparameters}: Material and scaling parameters
(T-dependent and T-independent) used in modeling.\label{tab:Materialparameters}
\end{itemize}
\pagebreak

\section*{Figure Captions}

\begin{figure}[h]
\includegraphics[height=2in,keepaspectratio]{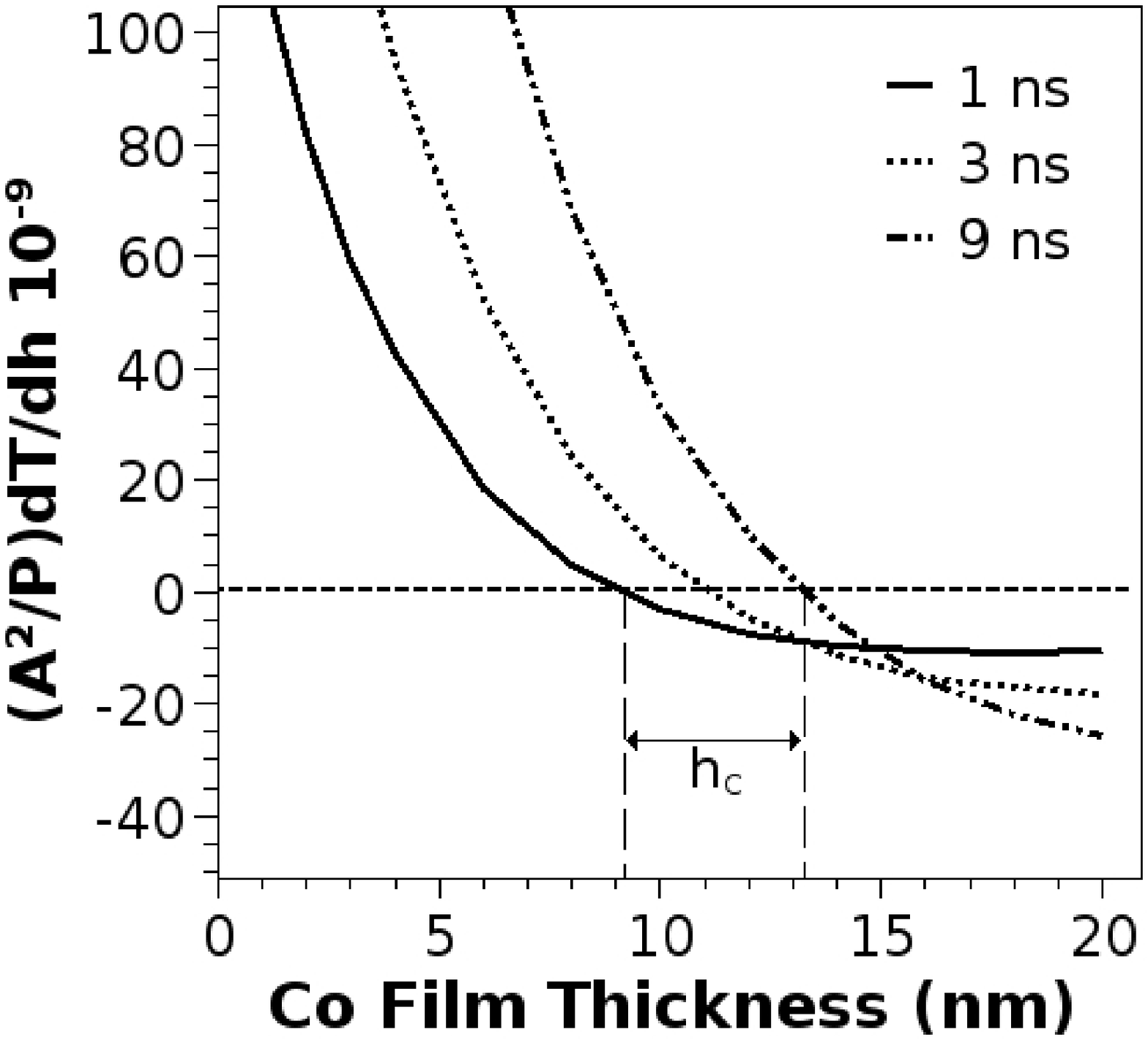}

\caption{\label{fig:dToverdh}}
\end{figure}

\begin{itemize}
\item Figure \ref{cap:Schematic}: (a) Schematic diagram for modeling of
laser heating of a metallic nanolayer on a semi-infinite insulating
substrate. (b) Plot of thickness-dependence of reflectivity R of Co/SiO$_{\text{2}}$
bilayer estimated from Ref. \cite{Heavens55}. \label{cap:Schematic}
\item Figure \ref{cap:Comparison-of-analytic-numeric}: Thermal behavior
of the Co film on SiO$_{\text{2}}$ irradiated by a pulsed laser.
(a) Time-dependent temperature obtained for the analytic and numerical
model using temperature independent parameters. The symbols are from
the analytical model while the lines are from the numerical calculations.
(b) Numerical results incorporating the phase change and T-dependent
parameters for various film thickness. (c) Laser energy density threshold
for melting $E_{m}$ of Co films on SiO$_{\text{2}}$. Comparison
of experimental measurement (solid circles) with calculations from
our analytical model (solid line) and numerical model (dashed line).
\label{cap:Comparison-of-analytic-numeric}
\item Figure \ref{fig:dToverdh}: The thermal models predict an intrinsic
thermal gradient $\partial T/\partial h$ whose magnitude and sign
was dependent on film thickness and time to melt during the ultrafast
heating of nanoscopic metal films. As shown in the figure a critical
range of thickness $h_{c}$ is predicted over which the $\partial T/\partial h$
changes sign, and this value is dependent on the time to melt. For
the thickness regime investigated in this work ($h\leq7\, nm$), the
$\partial T/\partial h$ dissipates in \emph{t} < $t_{p}$ and this
intrinsic effect is negligible. \label{fig:dtoverdheffect}
\item Figure \ref{cap:2.4nmpatterns}: Dewetting pattern evolution in a
$2\, nm$ Co film irradiated at $200\, mJ/cm^{\text{2}}$ as a function
of number of pulses $n$. The top row are SEM images while the bottom
row show corresponding power spectrum. (a) $n=10$; (b) $n=100$;
(c) $n=1000$; (d) $n=10500$. All the power spectra have an annular
structure indicating a band of spatial frequencies implying short
range spatial order. \label{cap:2.4nmpatterns}
\item Figure \ref{cap:2.4nmPatternvsE}: Dewetting pattern evolution in
a $2\, nm$ Co film irradiated as a function of laser energy density
for irradiation by 100 pulses. (a) $190\, mJ/cm^{\text{2}}$; (b)
$200\, mJ/cm^{\text{2}}$; (c) $220\, mJ/cm^{\text{2}}$ ; (d) $250\, mJ/cm^{\text{2}}$.
\label{cap:2.4nmPatternvsE}
\item Figure \ref{cap:3.5nmpatterns}: Dewetting pattern evolution in a
$4.4\, nm$ Co film irradiated at $93\, mJ/cm^{\text{2}}$ as a function
of number of pulses $n$. The top row are SEM images while the bottom
row show corresponding power spectrum. (a) $n=10$; (b) $n=100$;
(c) $n=1000$; (d) $n=10500$. Similar to the results for the $2\, nm$
film, all the power spectra have an annular structure indicating a
band of spatial frequencies implying short range spatial order. \label{cap:3.5nmpatterns}
\item Figure \ref{cap:3.5nmLvsnPS}: Quantitative analysis of the spatial
frequencies resulting from the short range order in the dewetting
patterns for the $4.4\, nm$ Co film. (a) The radial distribution
function $g(k)$ for the patterns of Fig. \ref{cap:3.5nmpatterns}
showing an initial increase as the holes expand and then a decrease
as the nanoparticles start to form. The vertical lines indicate peak
positions in the $g(k)$. (b) Behavior of the characteristic length
extracted from the peak in the $g(k)$ as a function of irradiation
time $n$ at various laser energies. The spacing of the final state
is independent of the laser energy. \label{cap:3.5nmLvsnPS}
\item Figure \ref{fig:Monomodal-size-distribution}: Monomodal size distribution
and average diameter of nanoparticles in the final robust state of
laser-induced dewetting. (a) An average diameter of $38\, nm$ results
for the $2\, nm$ Co film; (b) Average diameter of $207\, nm$ for
the $4.4\, nm$ Co film. \label{fig:Monomodal-size-distribution}
\item Figure \ref{fig: Spinodal-length-scale}: Analysis of the various
length scales in the final robust nanoparticle state as a function
of film thickness. (a) The average nearest-neighbour spacing $R$
of nanoparticles; (b) The average nanoparticle diameter $D$; (c)
The average areal density of nanoparticles $N$ in $\#/cm^{2}$. The
fits show exponents in good agreement with linear TFH theory. The
R$^{\text{2}}$ values in each graph indicate the correlation coefficient
of the least-squares fitting. \label{fig: Spinodal-length-scale}
\item Figure \ref{fig:Heating-cooling}: (a) Analytical result of the cooling
portion of thermal cycle for $3nm$ Co on SiO$_{\text{2}}$ showing
a fit using a second order exponential function. Cumulative results
of the analytical (closed symbols) and numerical model (open symbols)
using T-dependent materials parameters: (b) liquid lifetime vs. laser
energy density; and (c) peak temperature vs. laser energy density.
Closed symbols are analytical results while the open symbols are from
numerical calculations. \label{fig:Heating-cooling}
\item Figure \ref{cap:Role of Temperature}: Role of temperature on the
ripening rate of dewetting. (a) A linear dependence of the spinodal
ripening rate is predicted from the temperature dependence of the
Co liquid material parameters (Table. \ref{tab:Materialparameters}).
(b) The experimentally observed ripening rate $L/n$ for various film
thickness and energies. The symbols are experimental data while the
lines are linear fits. \label{cap:Role of Temperature} 
\end{itemize}
\pagebreak

\begin{table}[H]
\begin{tabular}{|>{\centering}p{0.75in}|p{1.7in}||p{3in}||c|}
\hline 
Quantity&
Description&
Value/Expression&
Reference\tabularnewline
\hline
\hline 
$\sigma=\frac{t_{p}}{2\cdot\sqrt{2\cdot ln(2)}}$&
standard deviation&
$3.822\, ns$&
\tabularnewline
\hline 
$R(h)$&
thickness-dependent reflectivity&
$R_{SiO_{2}}\leq R\leq R_{Co}$ &
\cite{OpticConst,Heavens55}\tabularnewline
\hline 
$\alpha_{m}^{-1}$&
absorbtivity&
$9.21\cdot10^{7}\, cm^{-1}$&
\cite{OpticConst}\tabularnewline
\hline
\hline 
$L$&
heat of transformation&
$274\,\frac{J}{g}\:$ for Co&
\cite{Linde92}\tabularnewline
\hline
\hline 
$T_{m}$&
melting temperature &
$1768$~K for Co&
\cite{Linde92}\tabularnewline
\hline
\hline 
$\rho_{m}$&
metal film density &
$(8.9\cdot(1-H(T-T_{m}))+7.75\cdot H(T-T_{m}))\cdot\frac{g}{cm^{3}}$&
\cite{Linde92}\tabularnewline
\hline
\hline 
$\rho_{s}$&
substrate density &
$2.2\,\frac{g}{cm^{3}}$&
\cite{Bansal86SiO2}\tabularnewline
\hline
\hline 
$(C_{eff}(T))_{m}$&
temperature dependent film heat capacity &
\parbox[t]{3in}{$(0.42\cdot(H(T)-H(T-T_{m}))$\\
$+0.59\cdot H(T-T_{m}))$$\frac{J}{g\cdot K}$}&
\cite{Linde92}\tabularnewline
\hline
\hline 
$(C_{eff})_{m}$&
temperature independent film heat capacity&
$0.42$$\frac{J}{g\cdot K}$&
\cite{Linde92}\tabularnewline
\hline
\hline 
$(C_{eff}(T))_{s}$&
temperature dependent substrate heat capacity &
$(0.931+(2.56\cdot10^{-4})\cdot T$$-0.240/T^{2})$$\frac{J}{g\cdot K}$&
\cite{Bansal86SiO2}\tabularnewline
\hline
\hline 
$(C_{eff})_{s}$&
temperature independent substrate heat capacity&
$0.937\,$$\frac{J}{g\cdot K}$&
\cite{Bansal86SiO2}\tabularnewline
\hline
\hline 
$k_{m}(T)$&
temperature dependent film thermal conductivity &
$(0.0912+264.65/T)\cdot(H(T)-H(T-1000))$\begin{eqnarray*}
+(-0.0016\cdot T+2.4402)\cdot(H(T-1000)-\end{eqnarray*}
$H(T-1400))$$\frac{W}{cmK}$&
\cite{Tyagi70,Montague79}\tabularnewline
\hline
\hline 
$k_{m}$&
temperature independent film thermal conductivity&
$1\,$$\frac{W}{cmK}$&
\cite{Linde92}\tabularnewline
\hline
\hline 
$k_{s}$&
substrate thermal conductivity&
$0.014\,$$\frac{W}{cmK}$&
\cite{Bansal86SiO2}\tabularnewline
\hline
\hline 
$\zeta$&
scaling constant&
0.76&
\tabularnewline
\hline
\hline 
$\epsilon$&
phase change tuning parameter&
$15$$K$&
\tabularnewline
\hline
\hline 
$T_{o}$&
room temperature&
$300$$K$&
\tabularnewline
\hline
\hline 
$\gamma(T=T_{mp})$&
surface tension&
$ $$1.88\, Jm^{-2}$&
\tabularnewline
\hline
\hline 
$\gamma_{T}$&
$d\gamma/dT$&
$-0.5\times10^{-3}Jm^{-2}K^{-1}$&
\tabularnewline
\hline
\hline 
$\eta(T)$&
viscosity&
$0.3272\times10^{-3}e^{\frac{0.3742\times10^{5}}{RT}}Pa-s$&
\tabularnewline
\hline
\hline 
$\eta_{T}$&
$d\eta/dT$&
$\frac{-4500.8}{T^{2}}\eta(T)Pa-s-K^{-1}$&
\tabularnewline
\hline
\end{tabular}

\caption{\label{tab:Materialparameters}}
\end{table}
\pagebreak

\pagebreak

\begin{figure}[H]
\begin{centering}\subfigure[]{\includegraphics[height=2in,keepaspectratio]{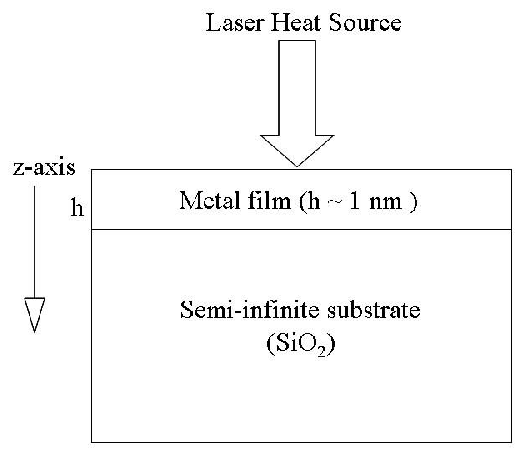}}\subfigure[]{\includegraphics[height=2in,keepaspectratio]{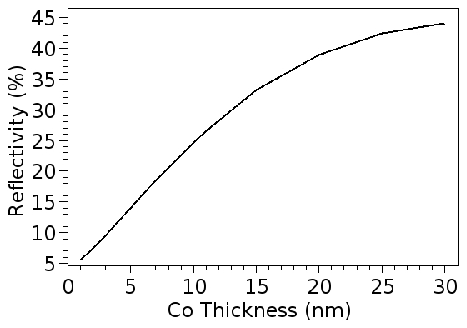}}\par\end{centering}

\caption{\label{cap:Schematic}}
\end{figure}
\pagebreak

\begin{figure}[h]
\begin{centering}\subfigure[]{\includegraphics[height=2.5in,keepaspectratio]{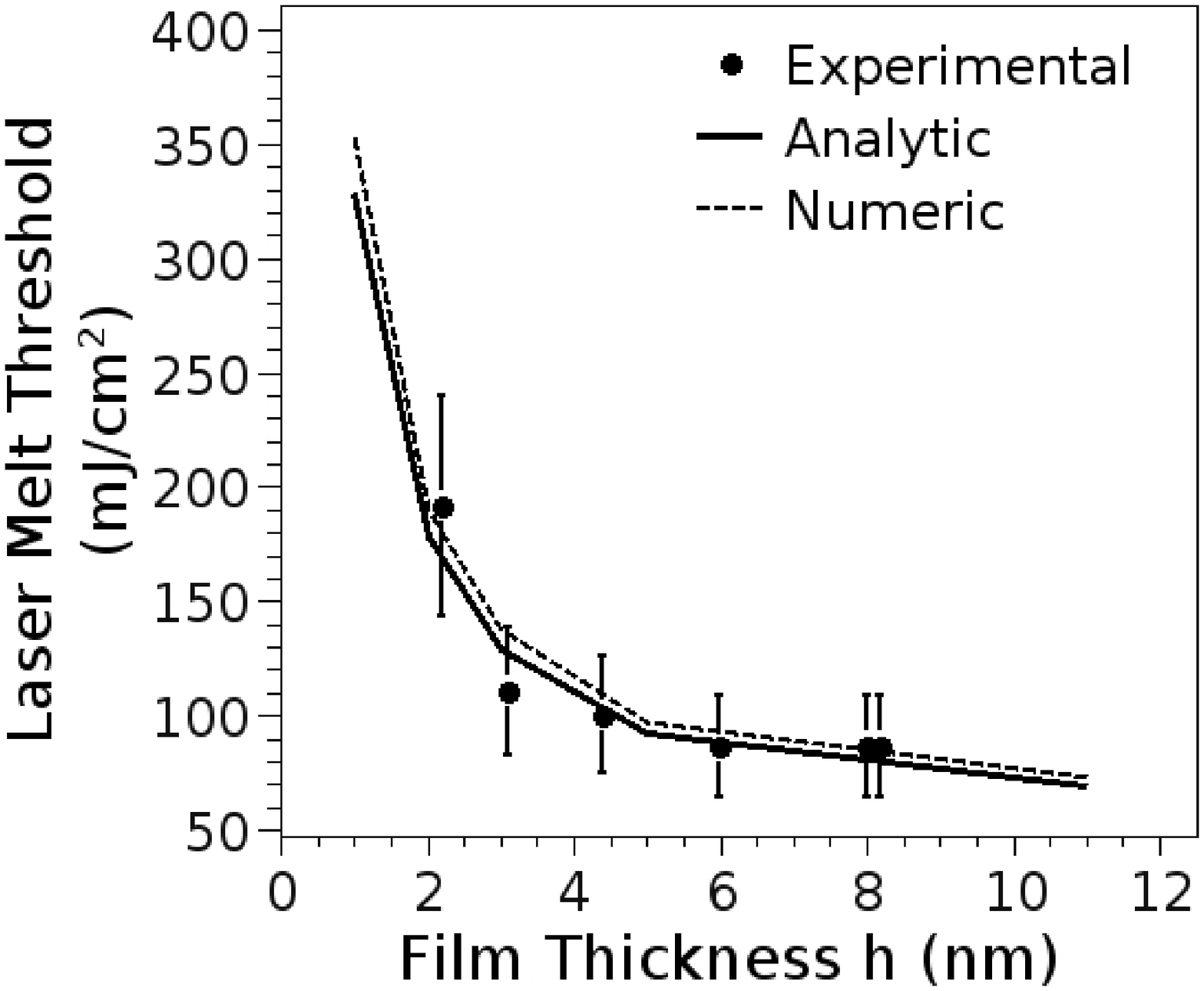}}\par\end{centering}

\begin{centering}\subfigure[]{\includegraphics[height=2.5in,keepaspectratio]{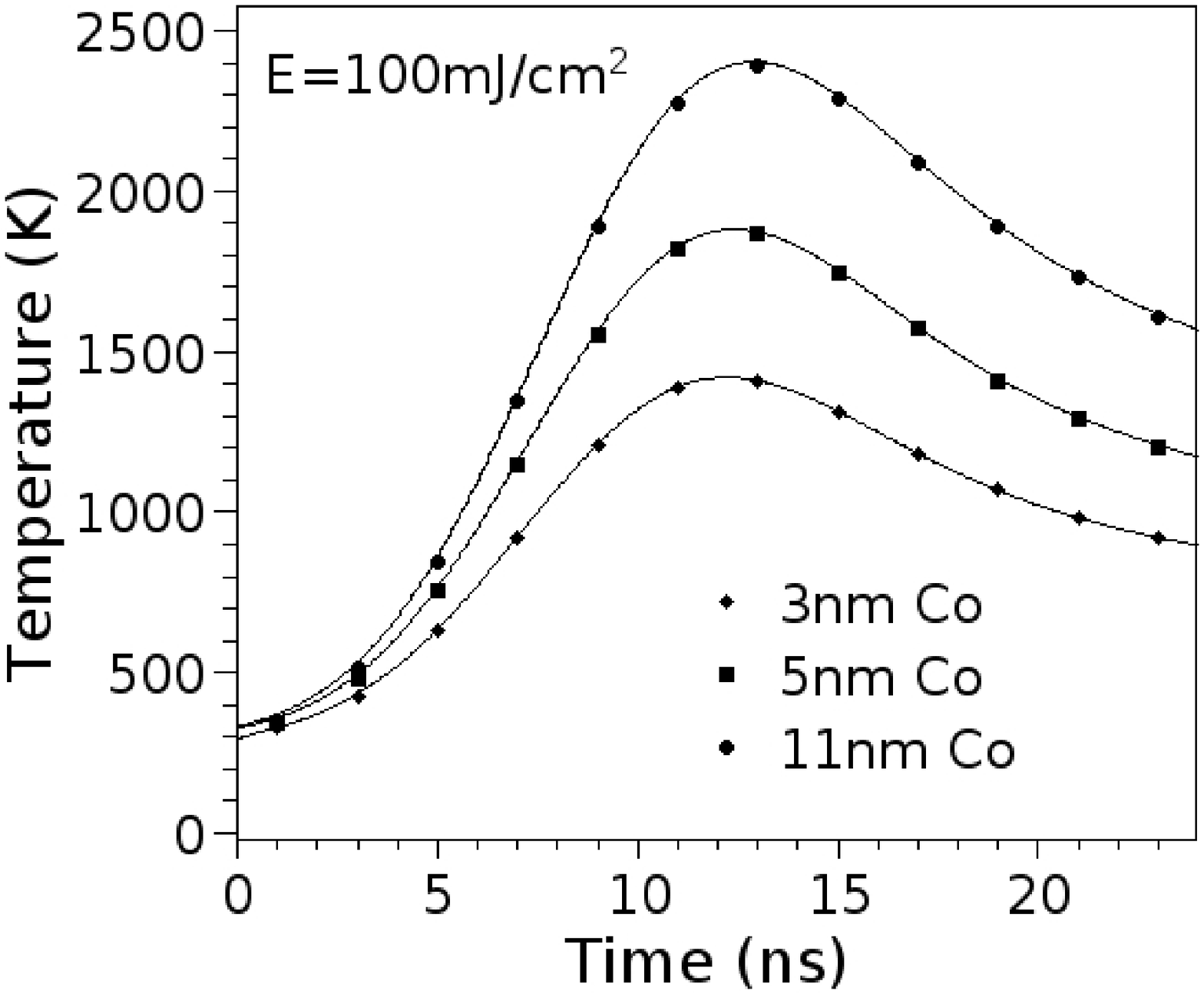}}\par\end{centering}

\begin{centering}\subfigure[]{\includegraphics[height=2.5in,keepaspectratio]{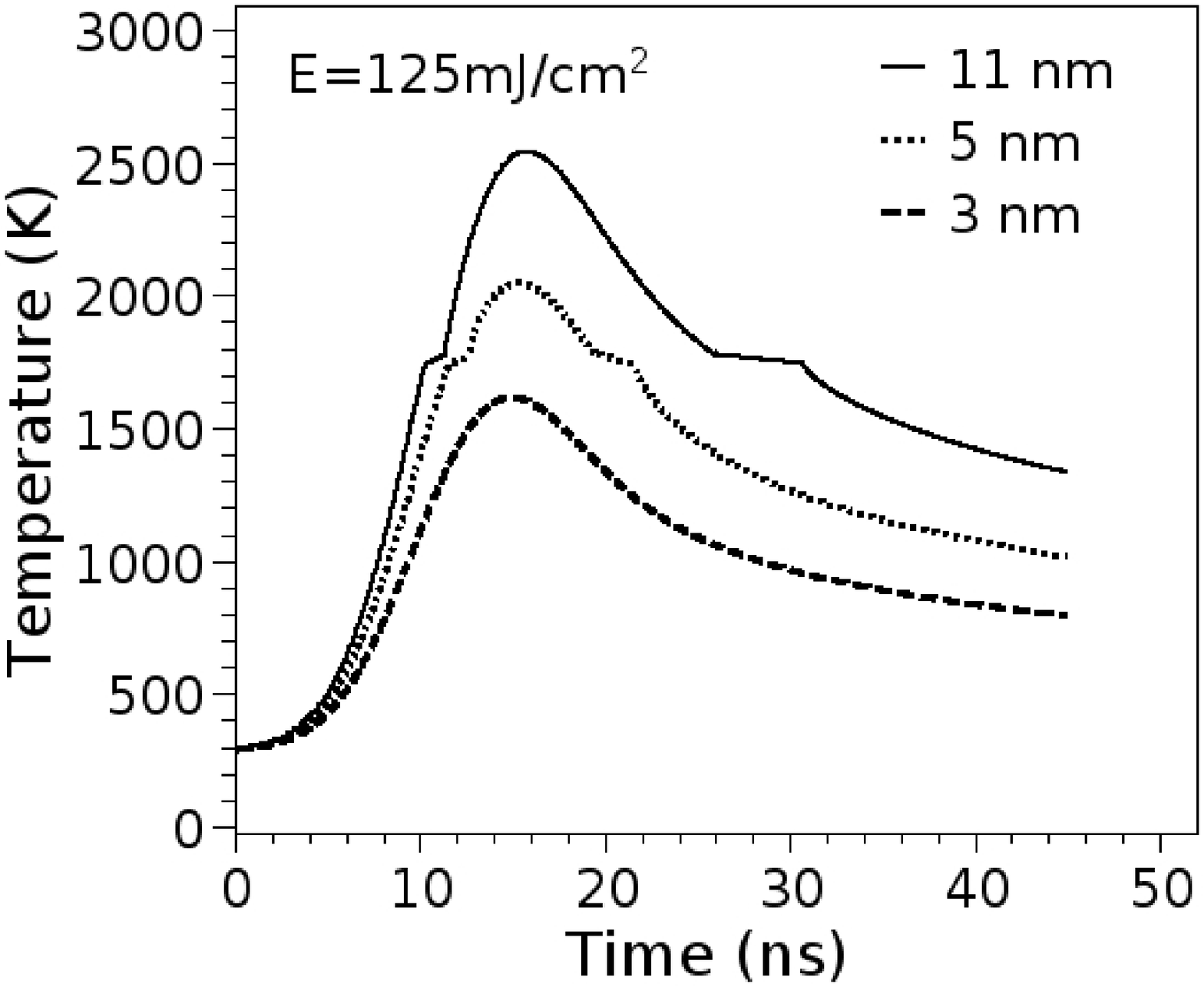}}\par\end{centering}

\caption{\label{cap:Comparison-of-analytic-numeric}}
\end{figure}
\pagebreak

\begin{figure}[h]
\begin{centering}\includegraphics[height=3in,keepaspectratio]{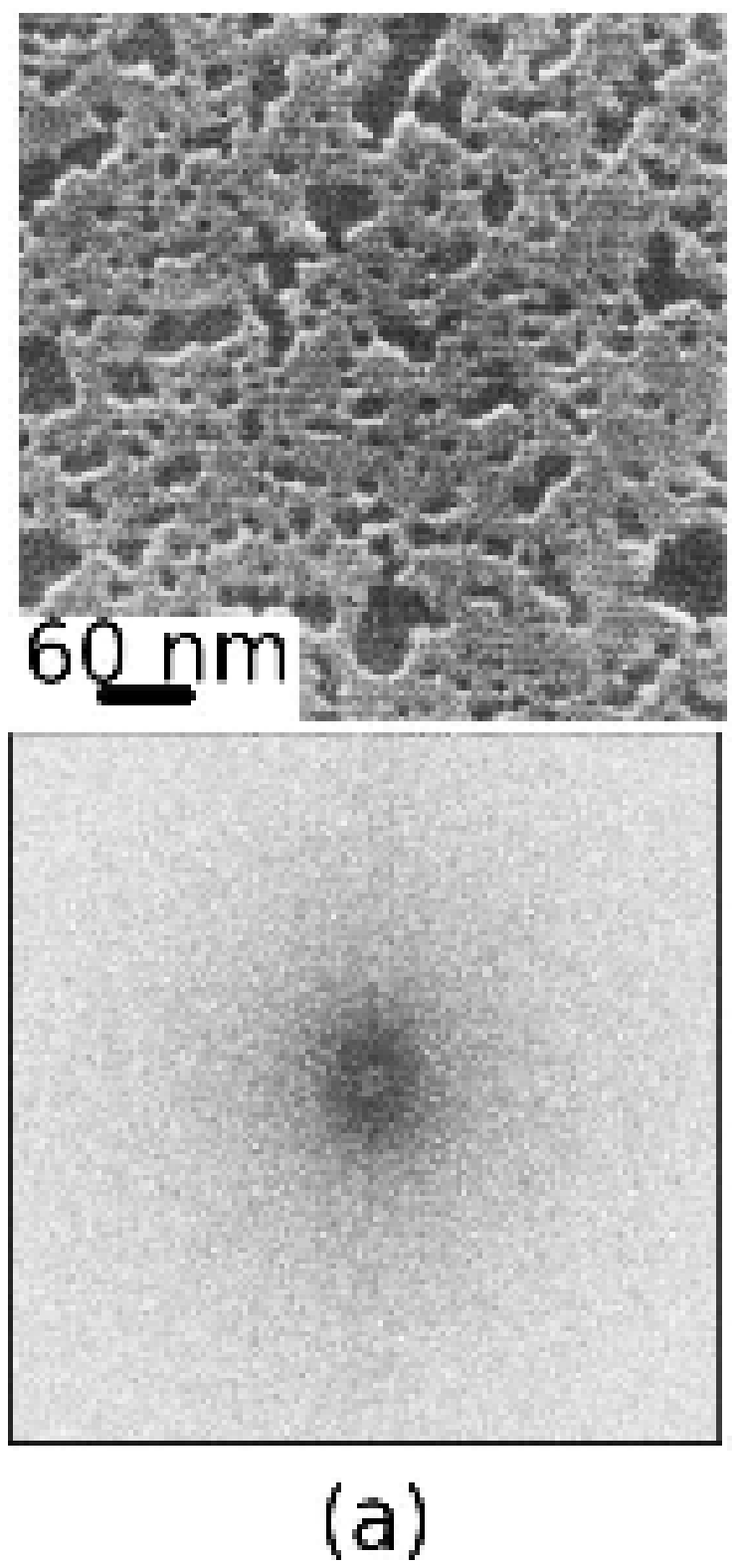}~\includegraphics[height=3in,keepaspectratio]{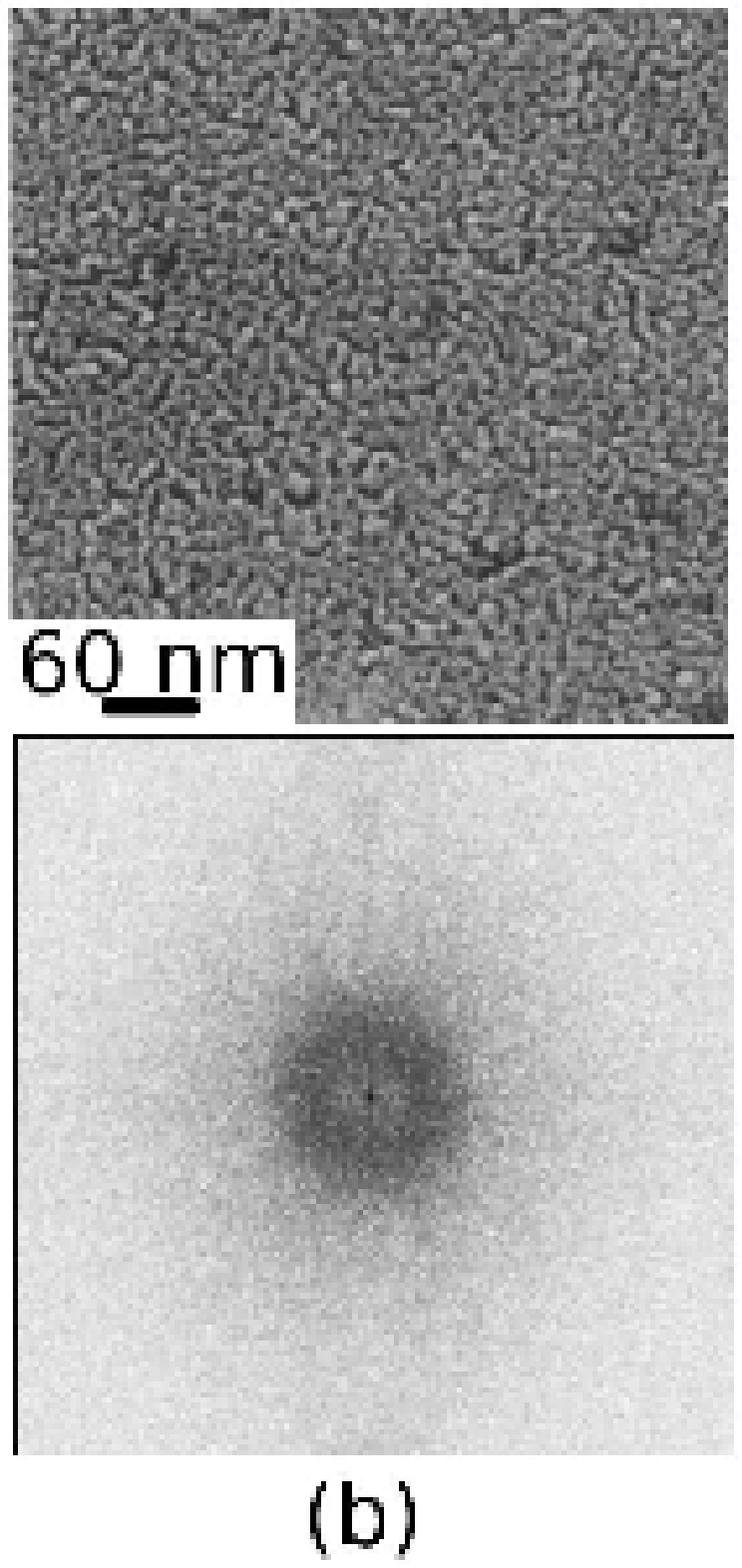}~\includegraphics[height=3in,keepaspectratio]{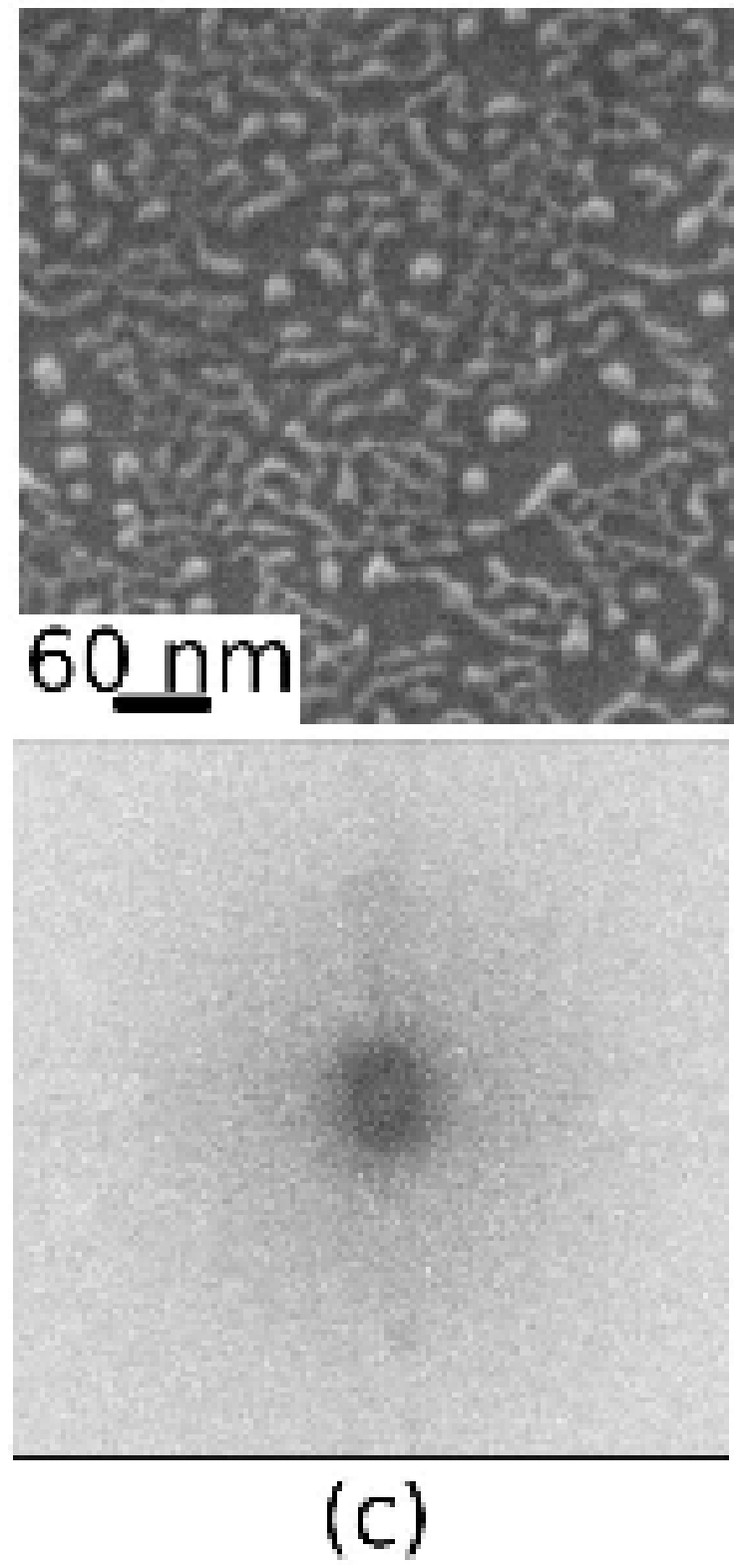}~\includegraphics[height=3in,keepaspectratio]{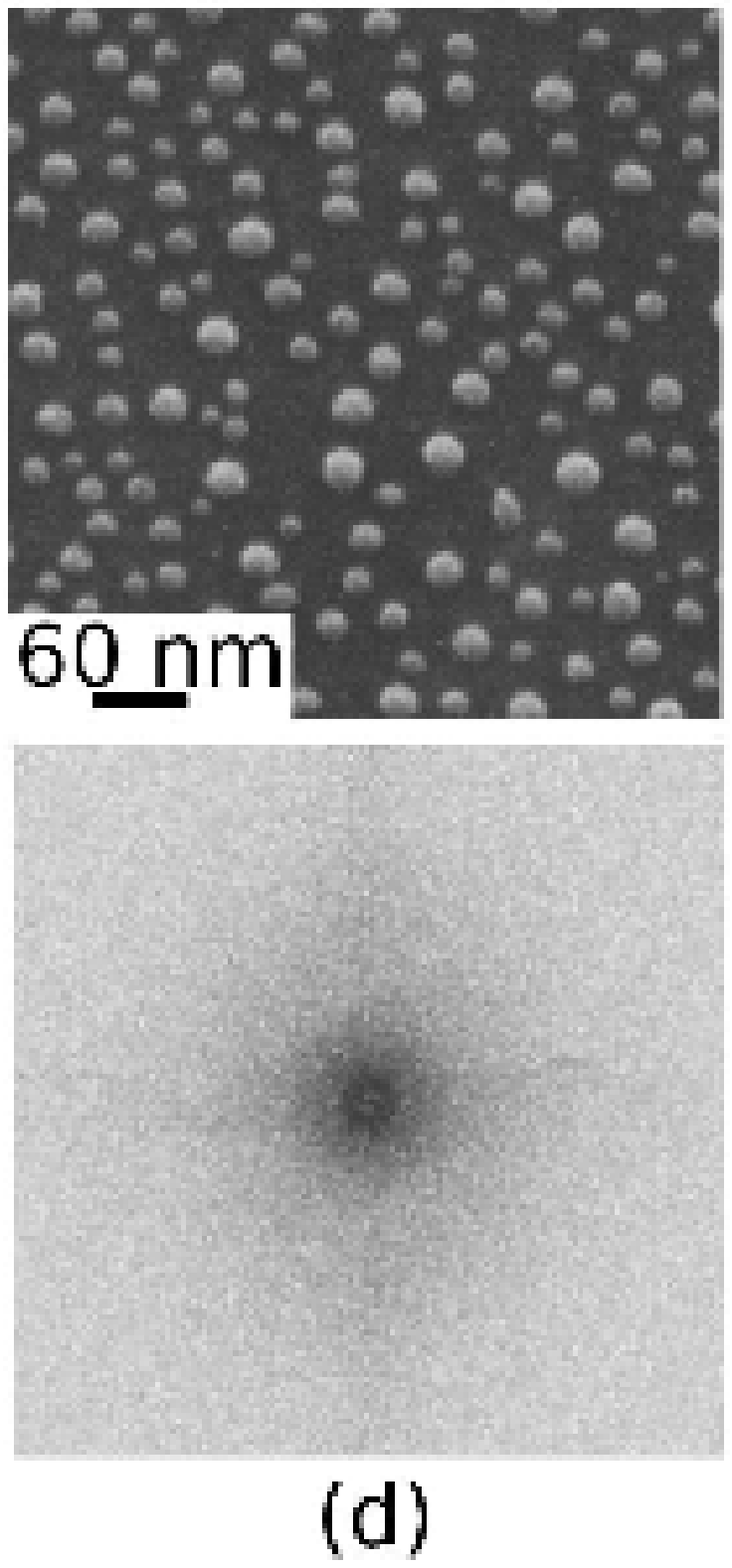}\par\end{centering}

\caption{\label{cap:2.4nmpatterns}}
\end{figure}

\begin{figure}[h]
\begin{centering}\subfigure[]{\includegraphics[width=1.45in,keepaspectratio]{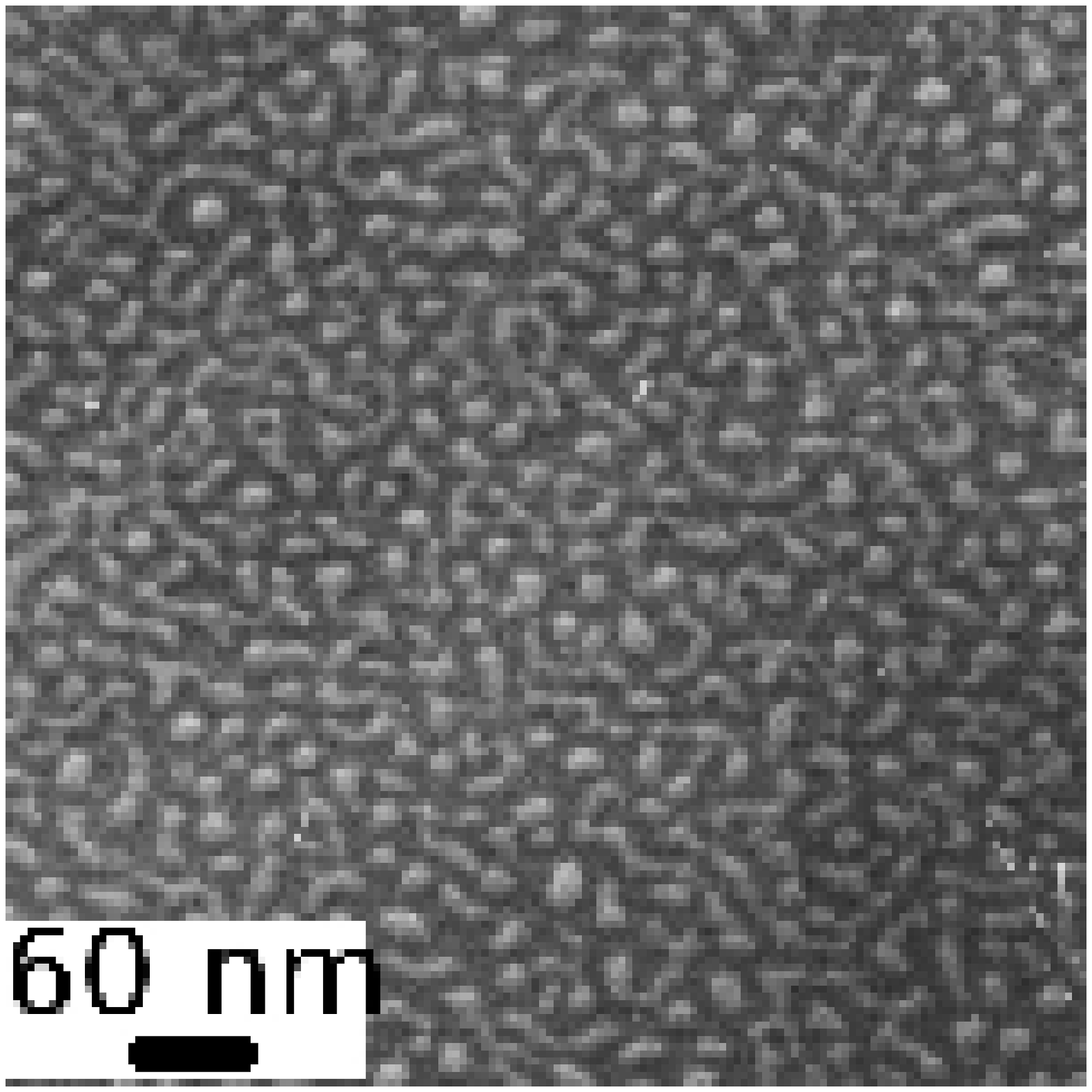}}~\subfigure[]{\includegraphics[width=1.45in,keepaspectratio]{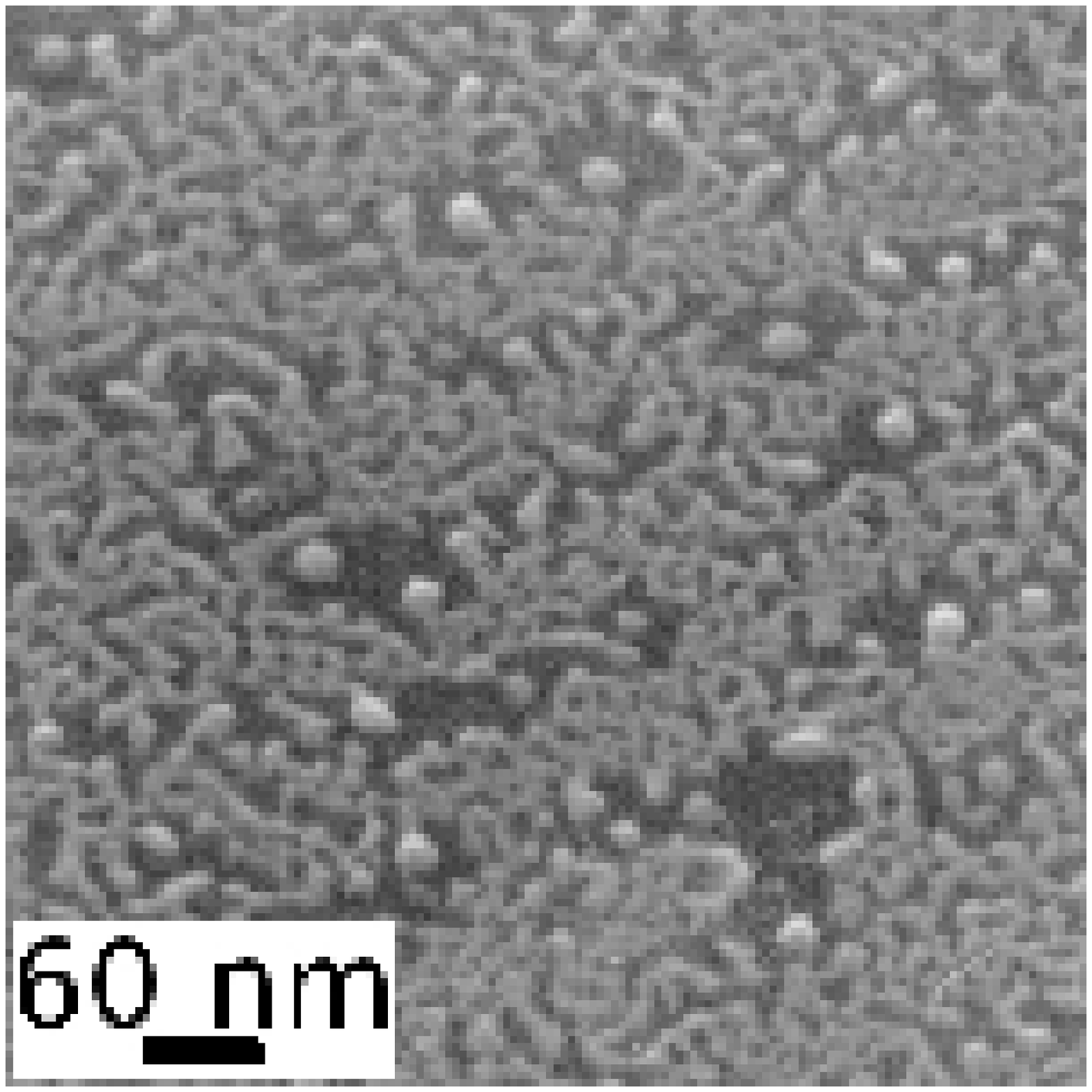}}~\subfigure[]{\includegraphics[width=1.45in,keepaspectratio]{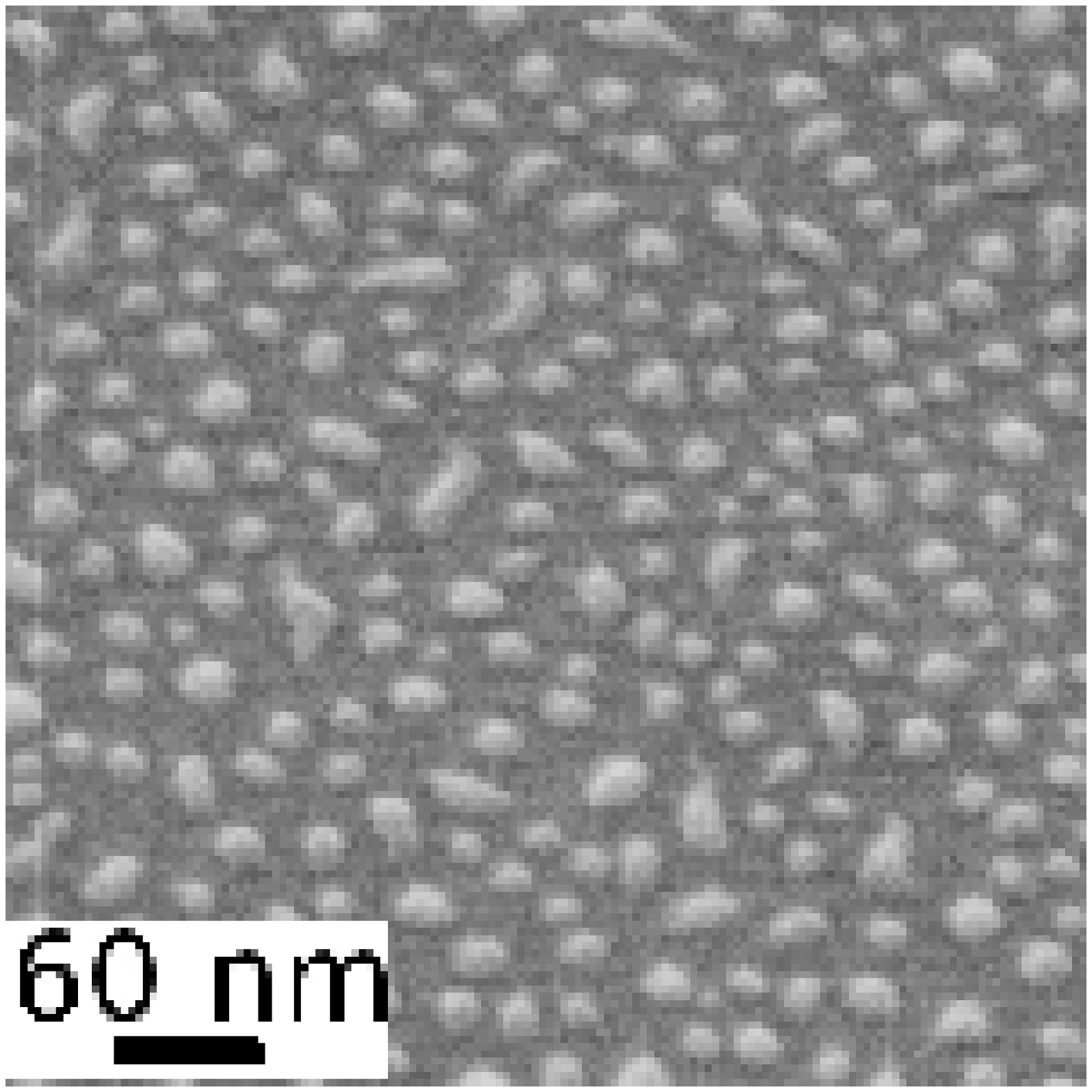}}~\subfigure[]{\includegraphics[width=1.45in,keepaspectratio]{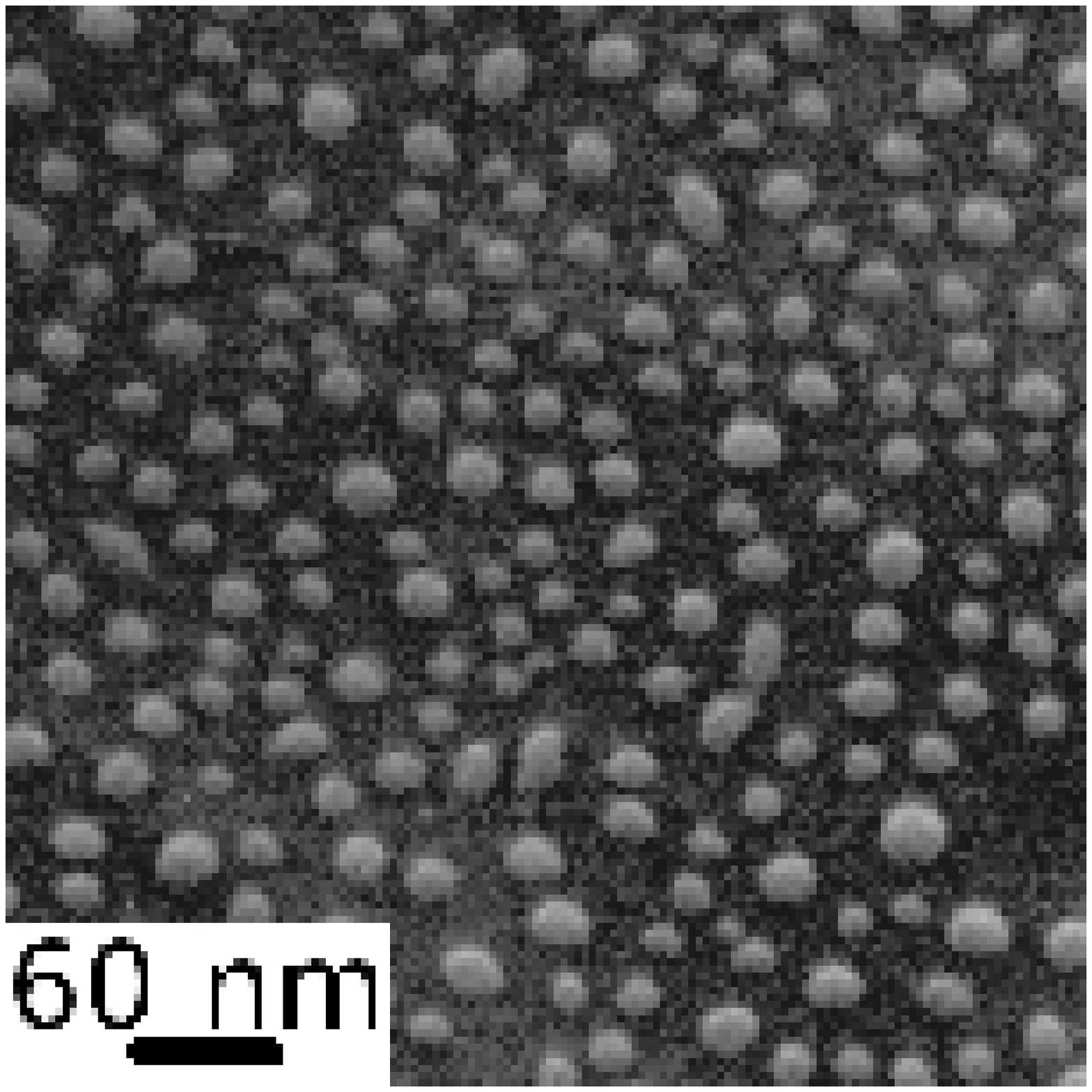}}\par\end{centering}

\caption{\label{cap:2.4nmPatternvsE}}
\end{figure}

\begin{figure}[h]
\begin{centering}\subfigure[]{\includegraphics[width=1.45in,keepaspectratio]{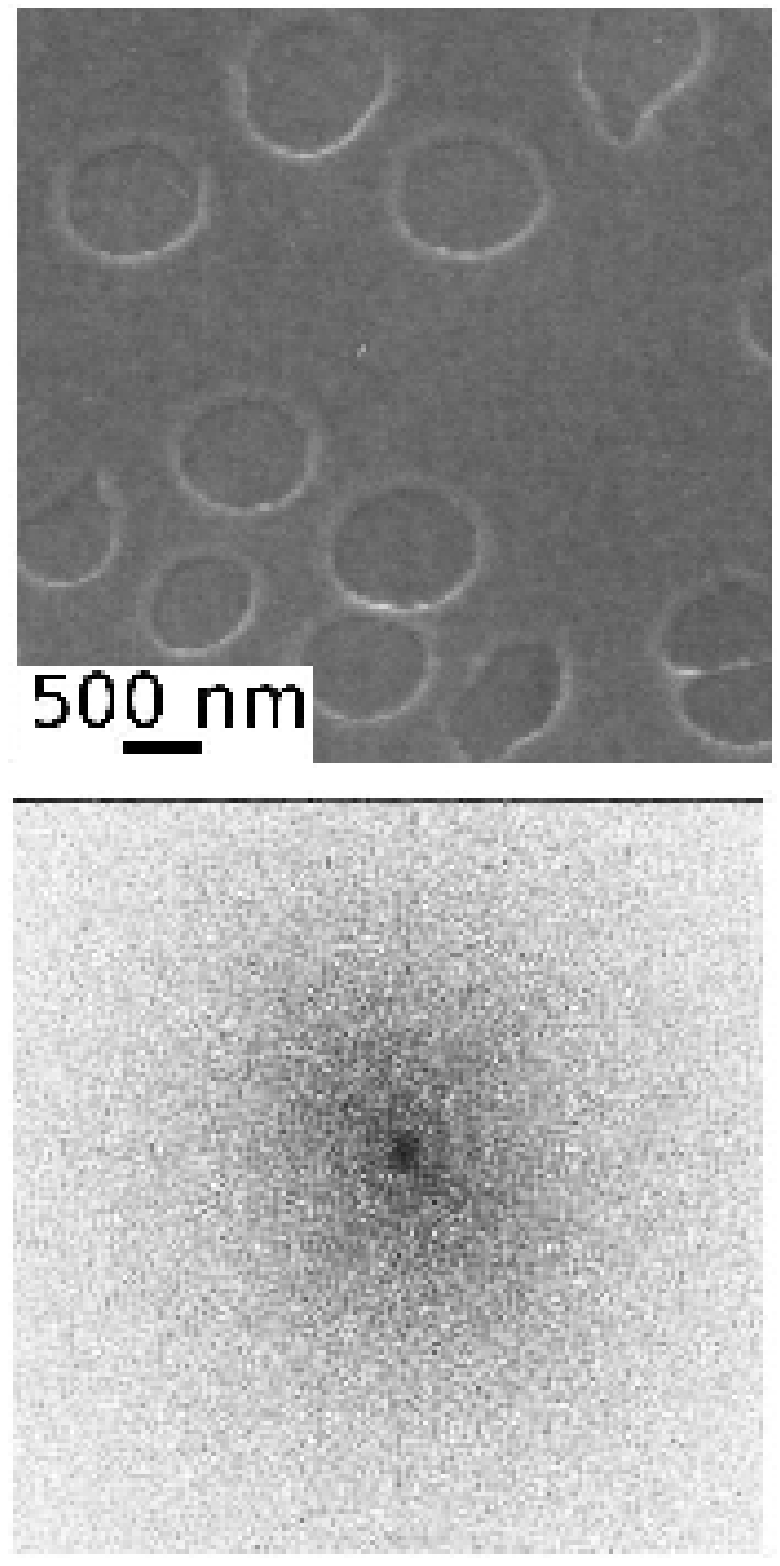}}~\subfigure[]{\includegraphics[width=1.45in,keepaspectratio]{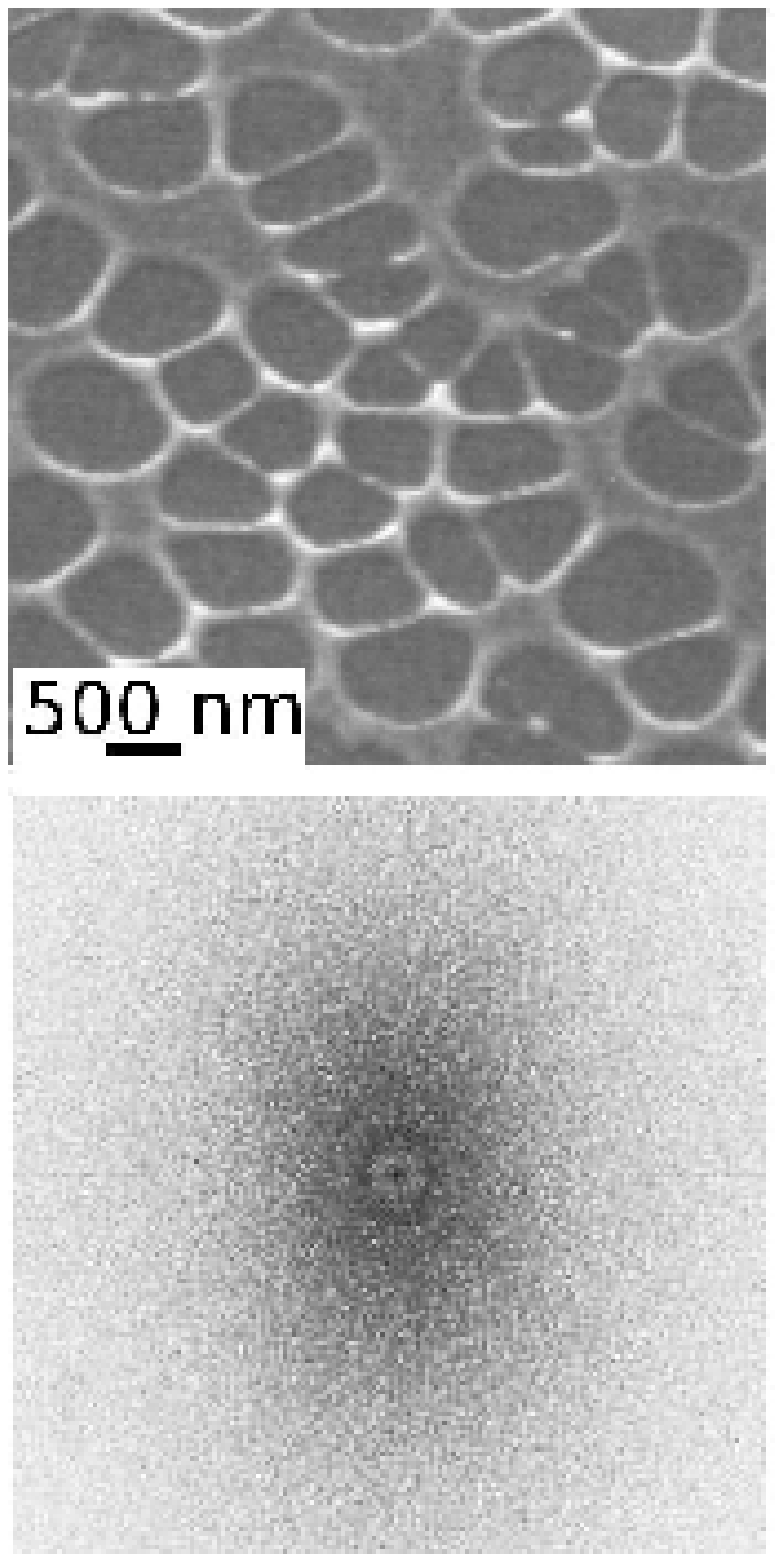}}~\subfigure[]{\includegraphics[width=1.45in,keepaspectratio]{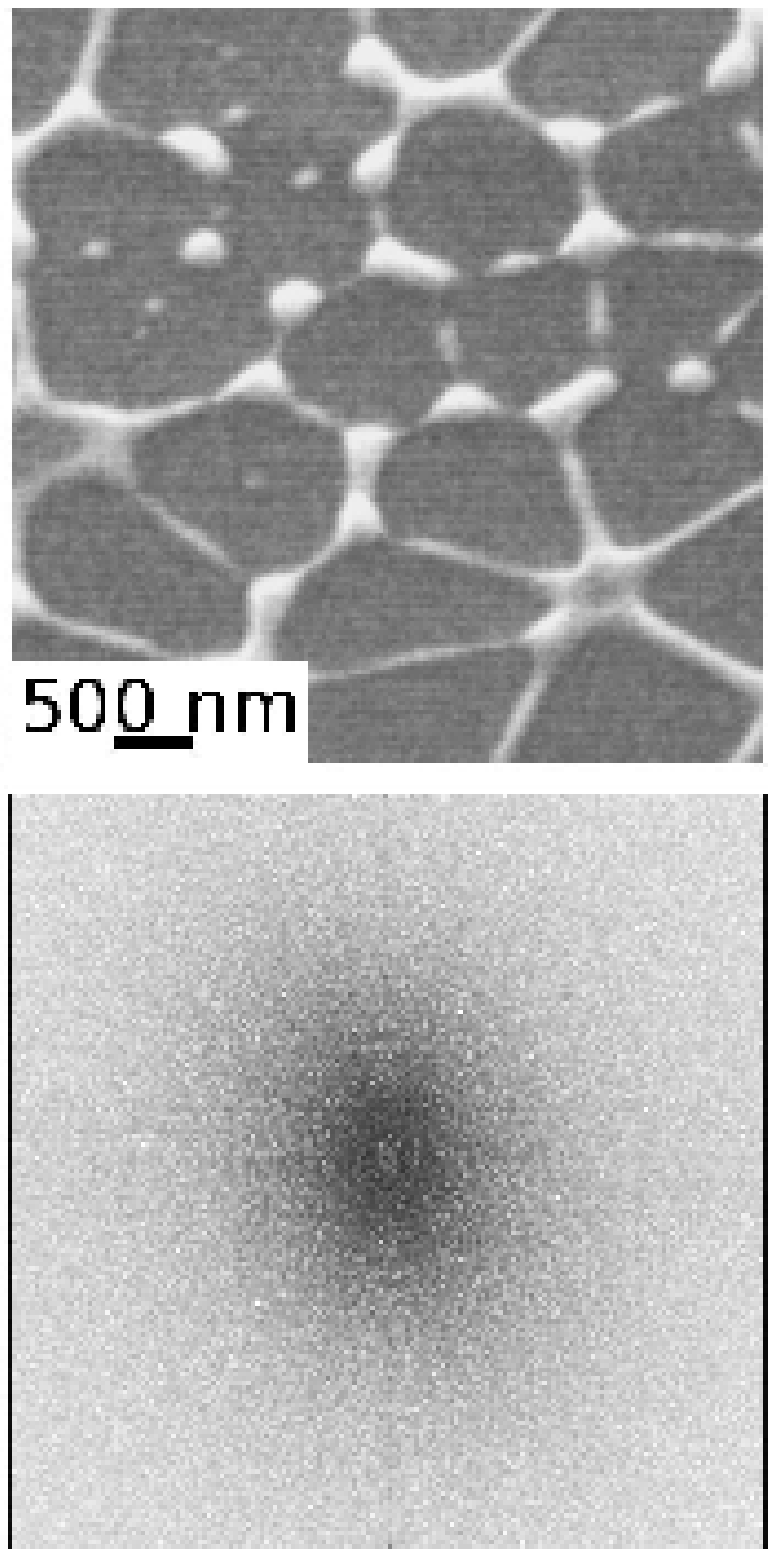}}~\subfigure[]{\includegraphics[width=1.45in,keepaspectratio]{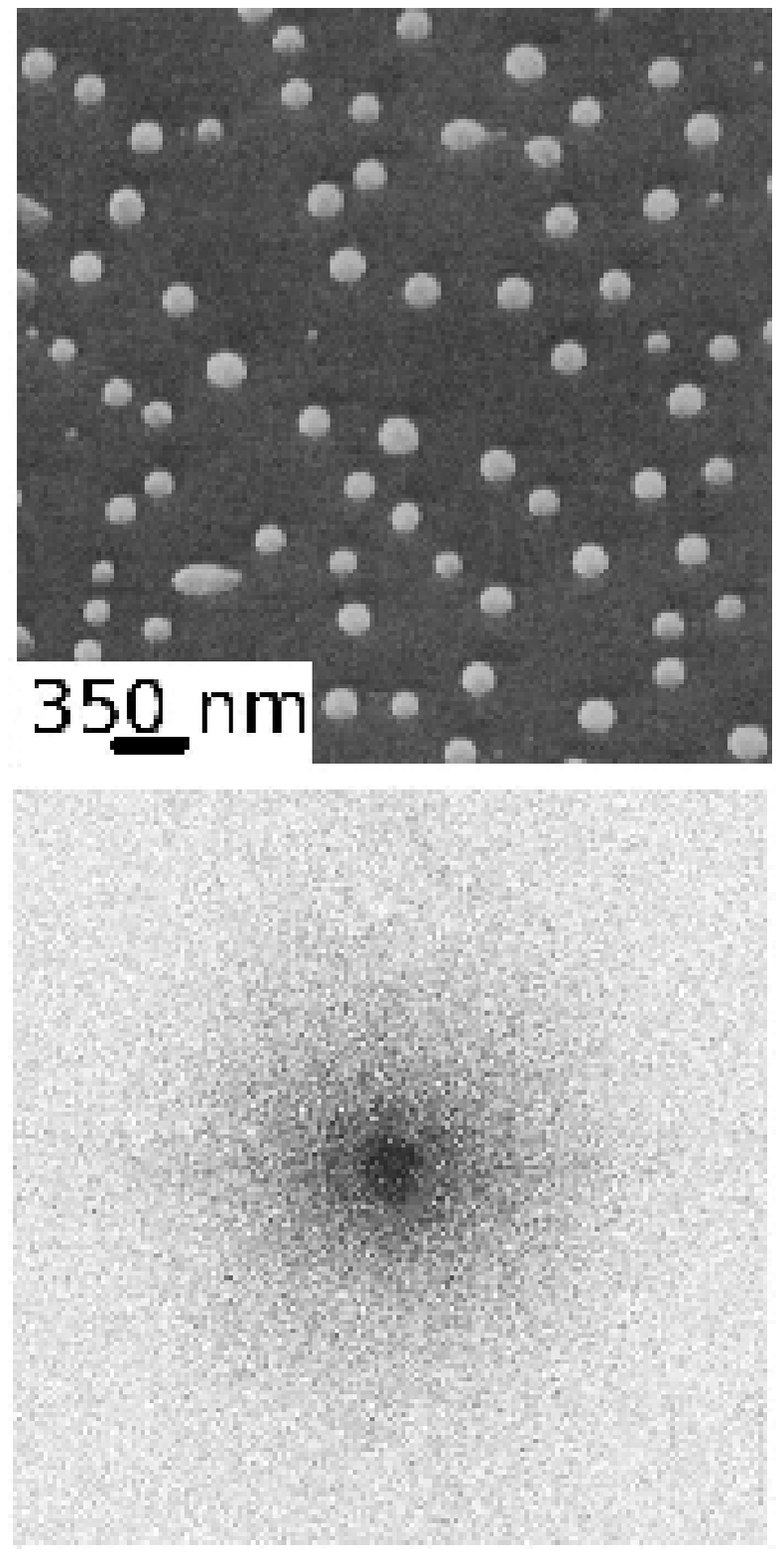}}\par\end{centering}

\caption{\label{cap:3.5nmpatterns}}
\end{figure}
\pagebreak

\begin{figure}[h]
\begin{centering}\subfigure[]{\includegraphics[height=2.5in,keepaspectratio]{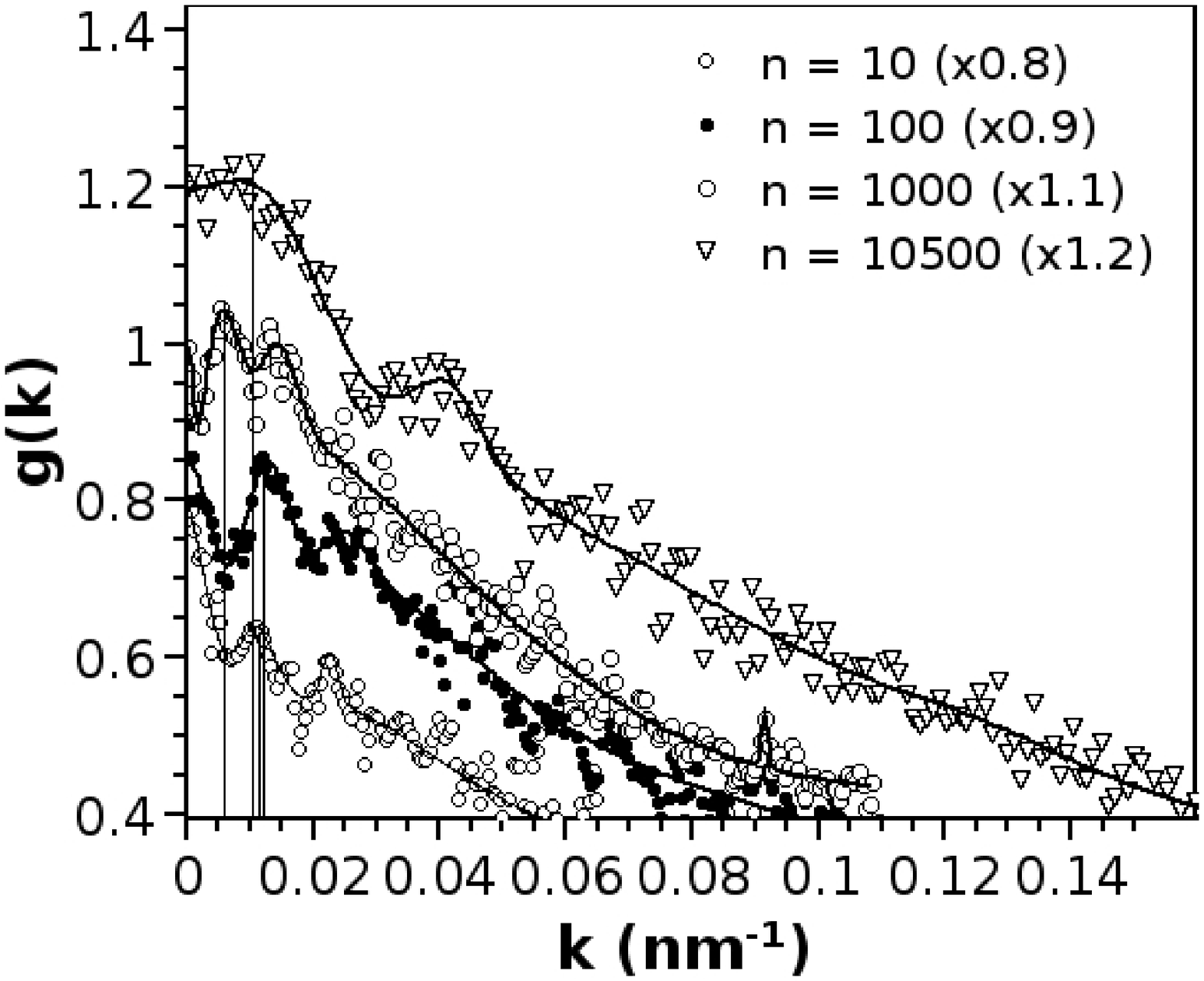}}~~\subfigure[]{\includegraphics[height=2.5in,keepaspectratio]{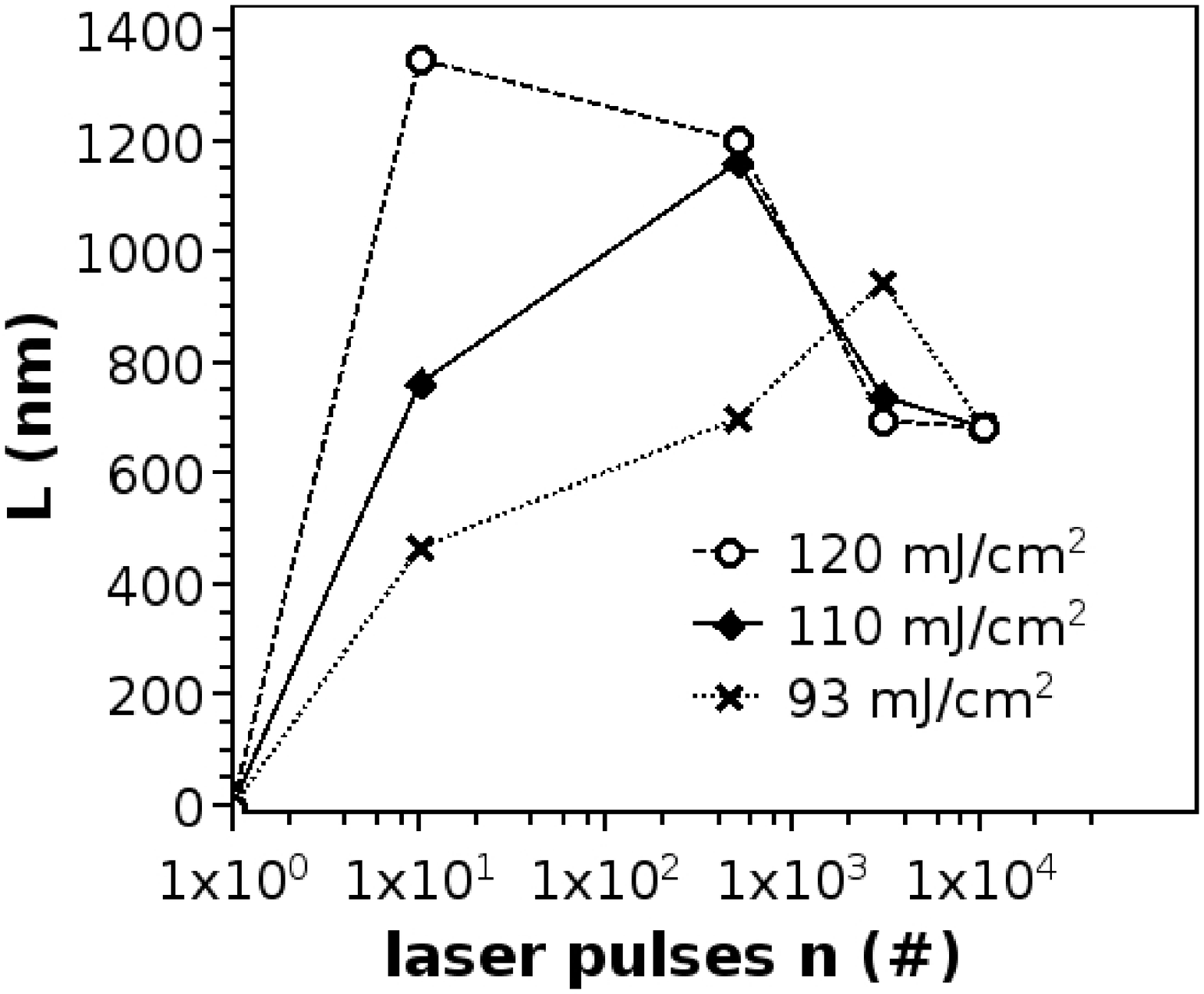}}\par\end{centering}

\caption{\label{cap:3.5nmLvsnPS}}
\end{figure}

\clearpage

\begin{figure}[h]
\begin{centering}\includegraphics[height=2.5in,keepaspectratio]{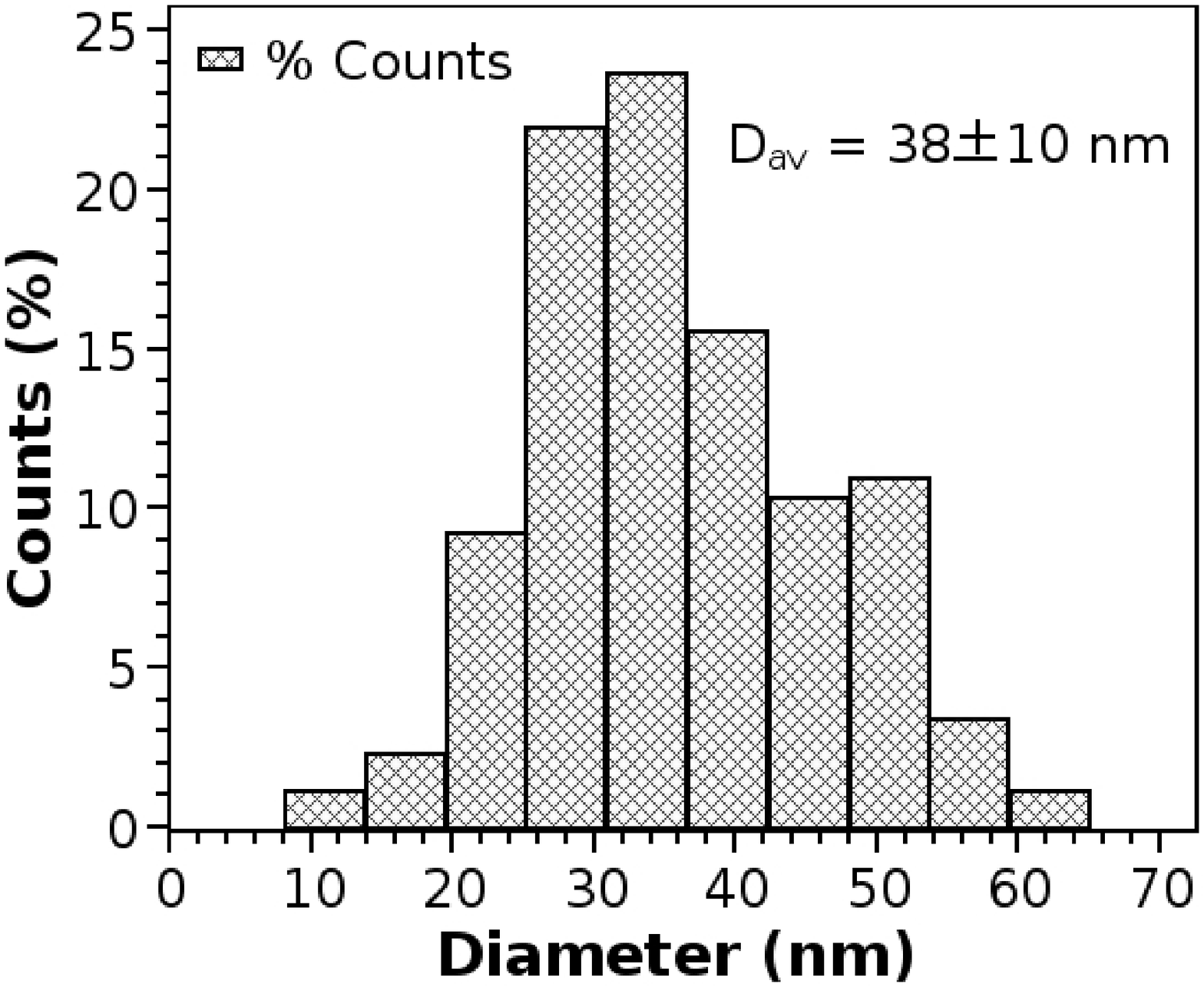}\includegraphics[height=2.5in,keepaspectratio]{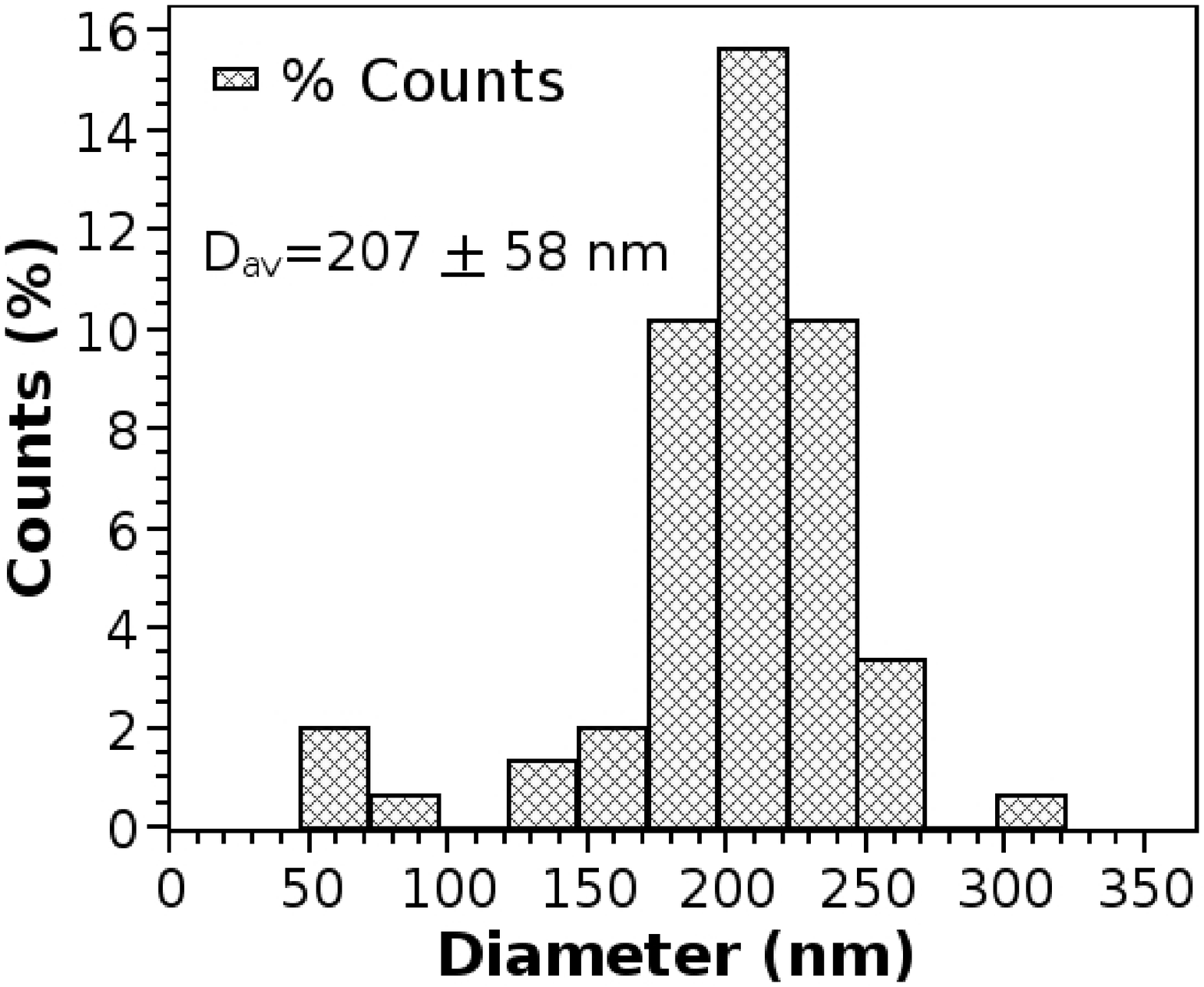}\par\end{centering}

\caption{\label{fig:Monomodal-size-distribution}}
\end{figure}

\pagebreak

\begin{figure}[h]
\begin{centering}\includegraphics[height=2.5in,keepaspectratio]{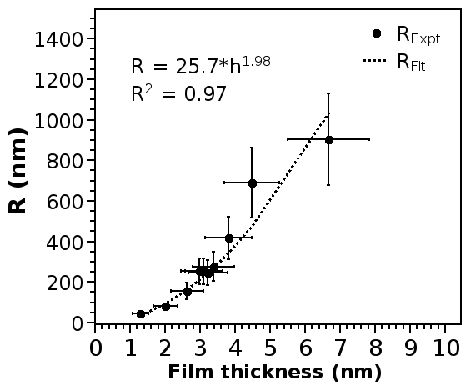}\par\end{centering}

\begin{centering}\includegraphics[height=2.5in,keepaspectratio]{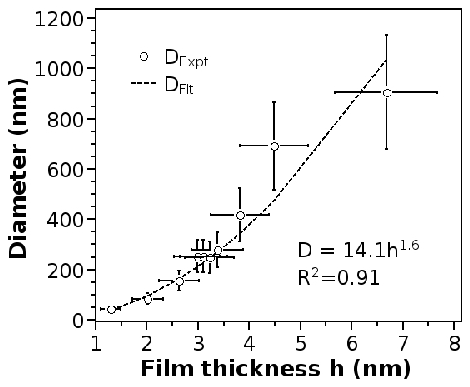}\par\end{centering}

\begin{centering}\includegraphics[height=2.5in,keepaspectratio]{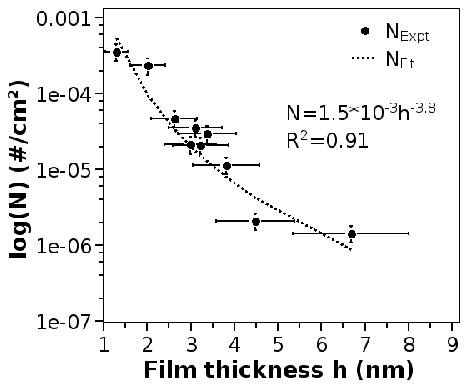}\par\end{centering}

\caption{\label{fig: Spinodal-length-scale}}
\end{figure}
\pagebreak

\begin{figure}[h]
\begin{centering}\subfigure[]{\includegraphics[height=2.5in,keepaspectratio]{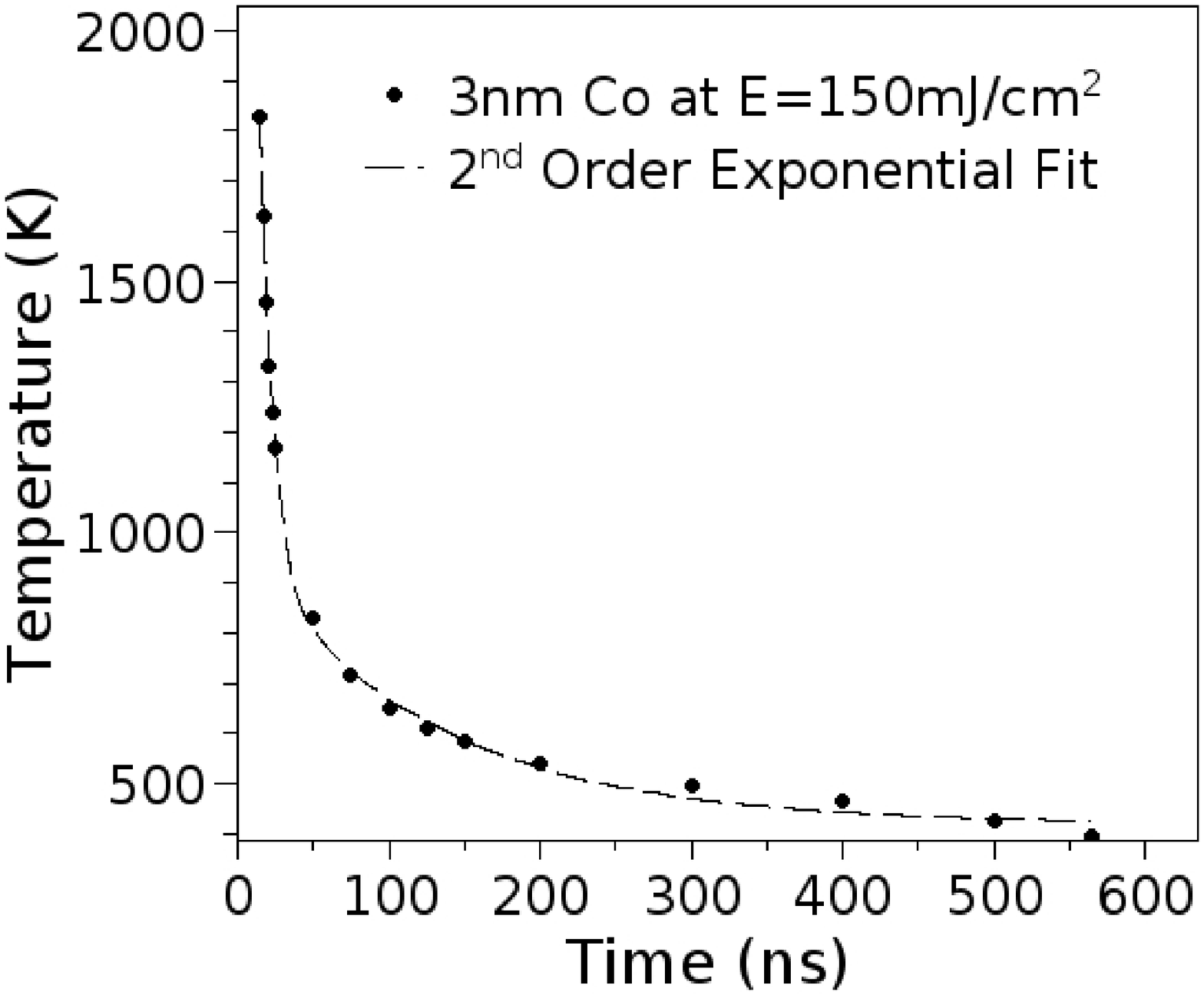}}\par\end{centering}

\begin{centering}\subfigure[]{\includegraphics[height=2.5in,keepaspectratio]{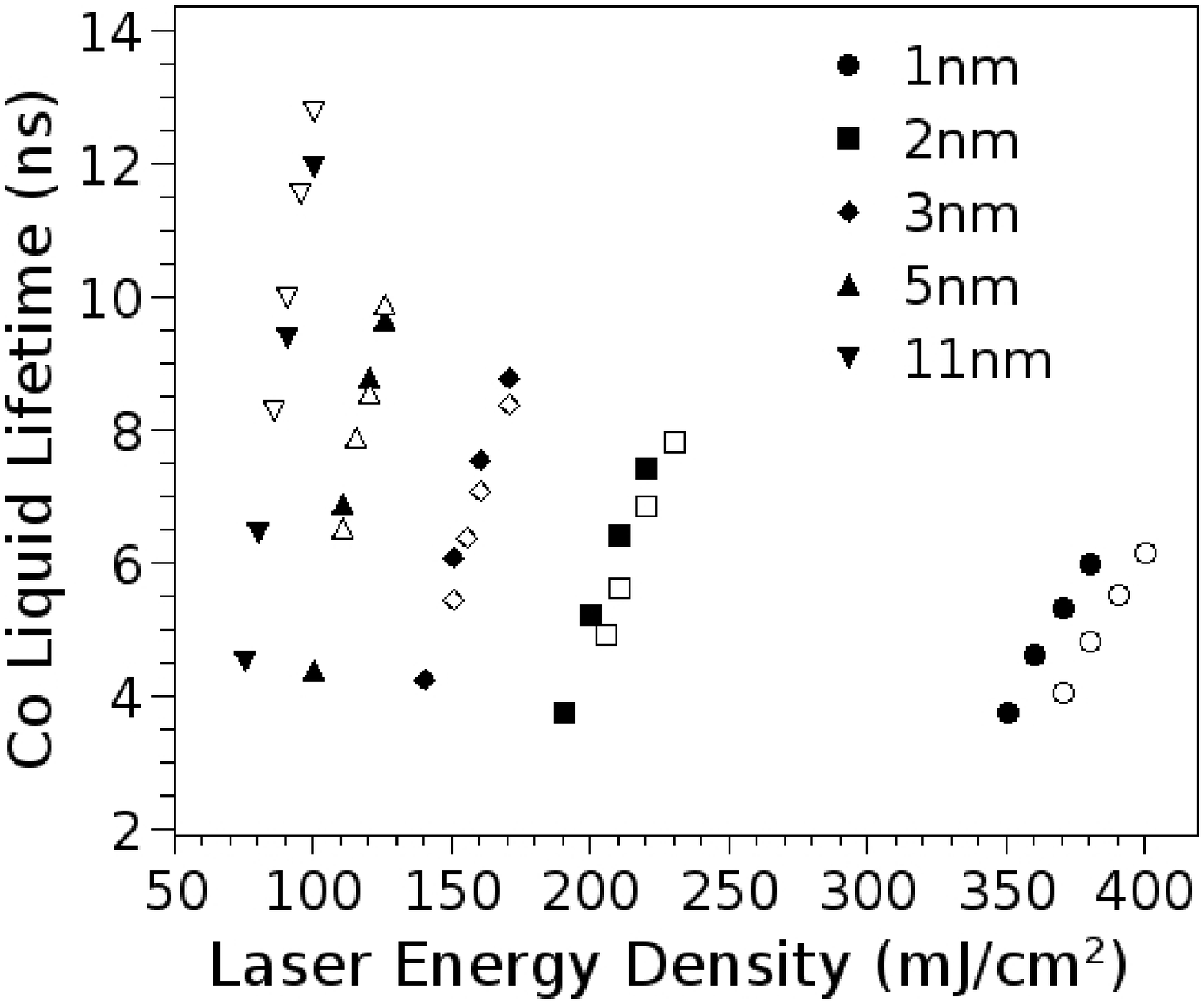}} \par\end{centering}

\begin{centering}~\subfigure[]{\includegraphics[height=2.5in,keepaspectratio]{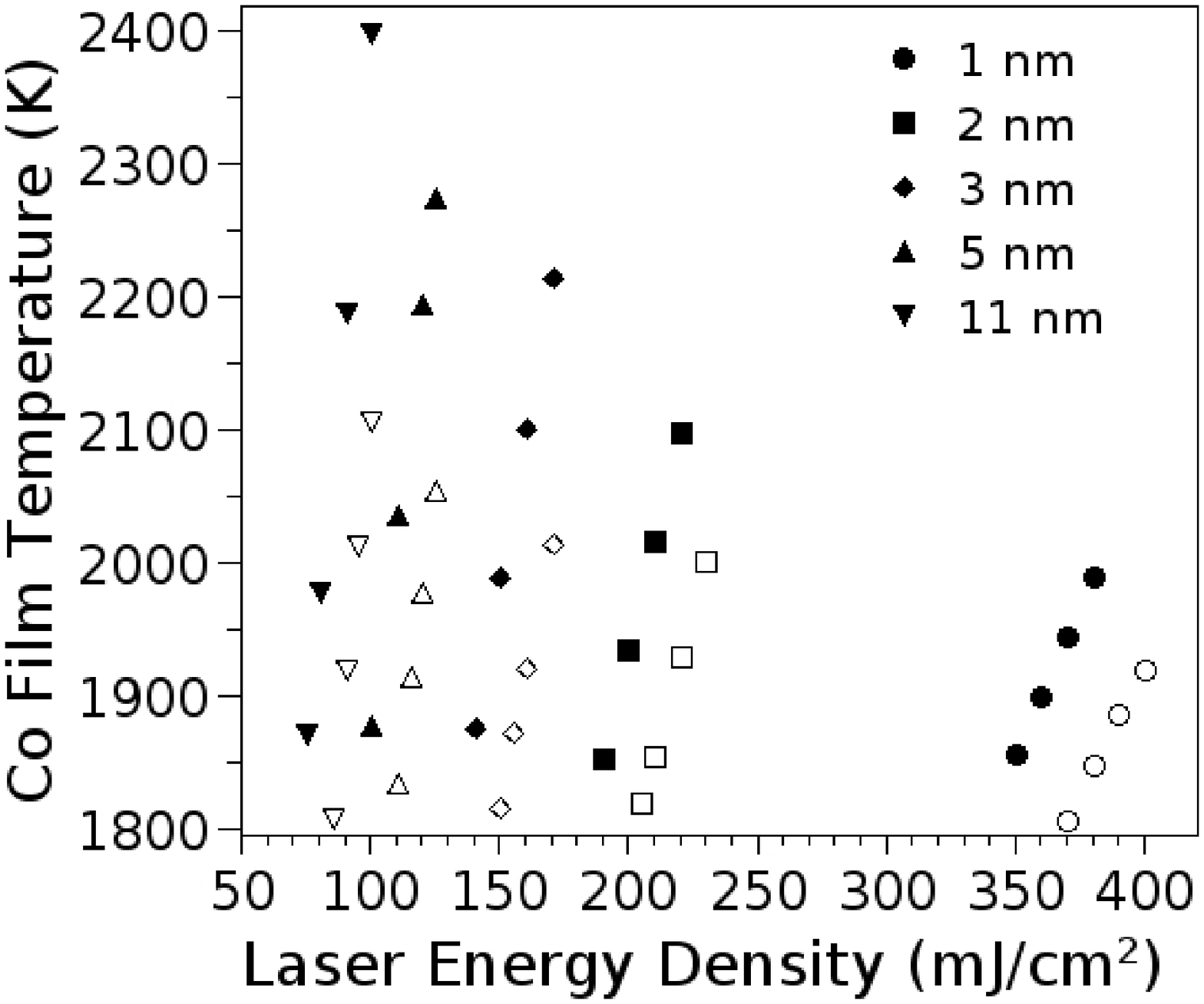}}\par\end{centering}

\caption{\label{fig:Heating-cooling}}
\end{figure}

\begin{figure}[h]
\begin{centering}\includegraphics[height=2.5in,keepaspectratio]{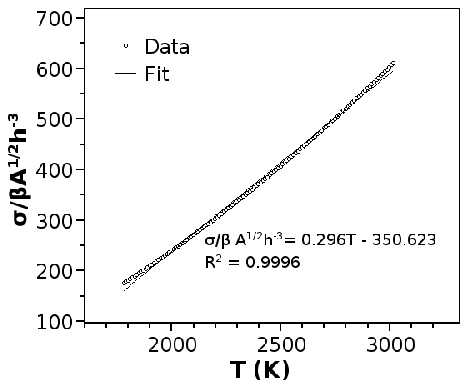}\par\end{centering}

\begin{centering}\includegraphics[height=2.5in,keepaspectratio]{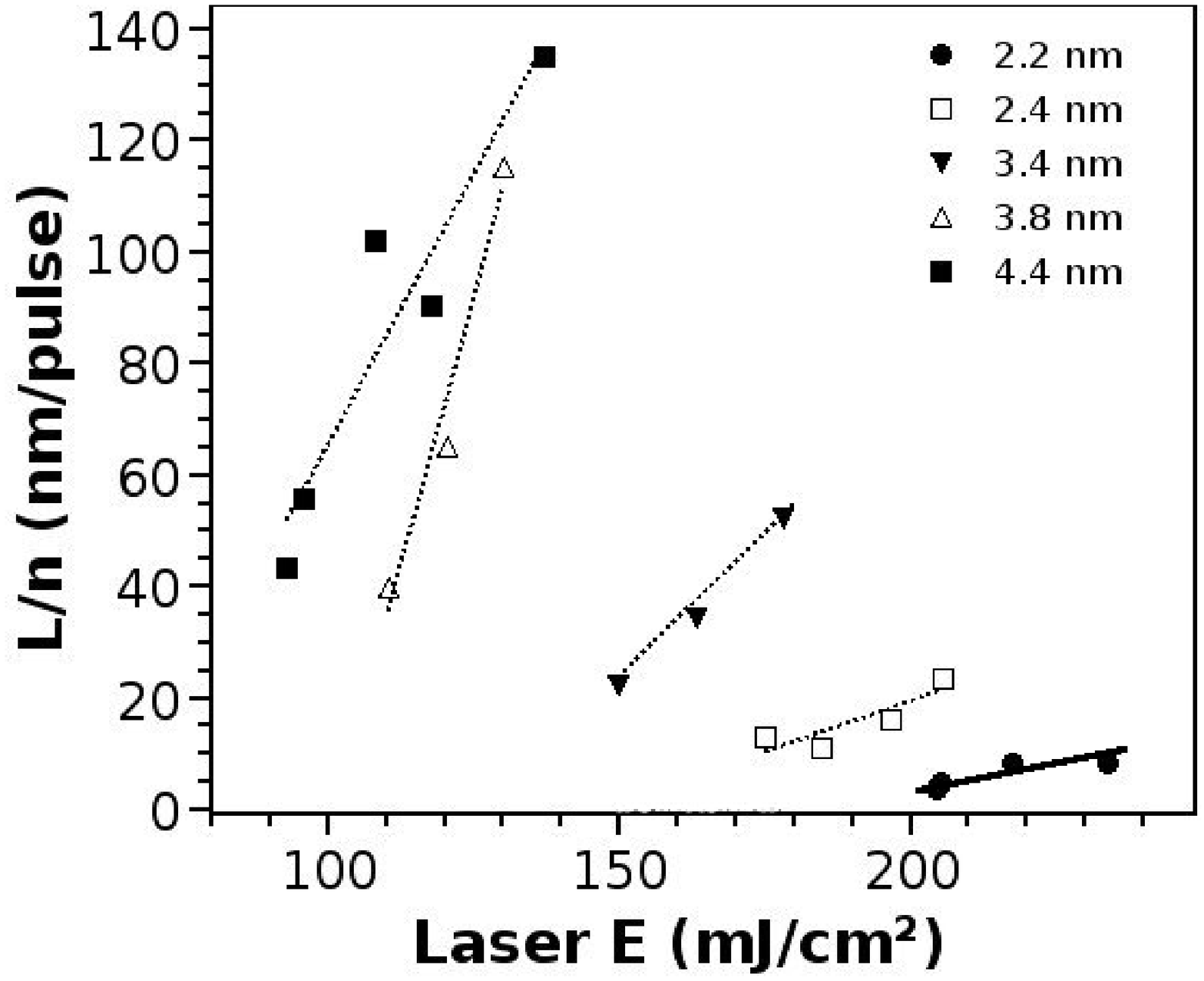}\par\end{centering}

\caption{\label{cap:Role of Temperature} }
\end{figure}

\end{document}